\documentclass[english,11pt]{article}\usepackage{lmodern}

\usepackage[T1]{fontenc}
\usepackage[latin9]{inputenc}
\usepackage[a4paper]{geometry}
\usepackage{graphicx}
\usepackage{bbm}
\usepackage{subfigure}
\usepackage{setspace}
\usepackage{fancyvrb}
\usepackage{appendix}
\usepackage{amssymb}

\usepackage{amsmath}
\usepackage{dcolumn}
\usepackage{lscape}
\usepackage{float}
\usepackage{tabulary}
\numberwithin{equation}{section}
\usepackage[unicode=true, pdfusetitle,
 bookmarks=true,bookmarksnumbered=false,bookmarksopen=false,
 breaklinks=true,pdfborder={0 0 0},backref=false,colorlinks=false]
 {hyperref}

\title{bayes}
\author{RM}

\begin{document}
\title{Understanding the Impact of Microcredit Expansions: A Bayesian Hierarchical Analysis of 7 Randomised Experiments \\
\vskip 14pt
\large WORKING PAPER\\}
\author{Rachael Meager\footnote{Massachusetts Institute of Technology (Graduate Student). Contact: rmeager@mit.edu}\; \thanks{I thank Esther Duflo, Abhijit Banerjee, Anna Mikusheva, Rob Townsend, Jeff Harris, Victor Chernozhukov, Andrew Gelman, Ben Olken, Jerry Hausman, Lars Hansen, Shira Mitchell, Kirill Boursayak, Cory Smith, Jonathan Huggins, Ryan Giordano, Tamara Broderick, Arianna Ornaghi, Greg Howard, Nick Hagerty, John Firth, Jack Liebersohn, Peter Hull, Matt Lowe, Yaroslav Mukhin, Tetsuya Kaji, Xiao Yu Wang, Aaron Pancost, and the participants of NEUDC 2015, The Chicago-MIT student conference 2016, the MIT Economic Development Lunch Seminar, MIT Econometrics Lunch Seminar and Yale PF/Labor Lunch Seminar for their suggestions, critiques, and advice. I also thank the authors of the 7 studies that I use in my analysis, and the journals in which they were published, for making their data and code public. All remaining mistakes are my own. This is a working paper, so please send critiques and corrections to rmeager@mit.edu}}
\date{\today}
\maketitle

\begin{abstract}
Bayesian hierarchical models are a methodology for aggregation and synthesis of data from heterogeneous settings, used widely in statistics and other disciplines. I apply this framework to the evidence from 7 randomized experiments of expanding access to microcredit to assess the general impact of the intervention on household outcomes and the heterogeneity in this impact across sites. The results suggest that the effect of microcredit is likely to be positive but small relative to control group average levels, and the possibility of a negative impact cannot be ruled out. By contrast, common meta-analytic methods that pool all the data without assessing the heterogeneity misleadingly produce ``statistically significant'' results in 2 of the 6 household outcomes. Standard pooling metrics for the studies indicate on average 60\% pooling on the treatment effects, suggesting that the site-specific effects are reasonably externally valid, and thus informative for each other and for the general case. The cross-study heterogeneity is almost entirely generated by heterogeneous effects for the 27\% households who previously operated businesses before microcredit expansion, although this group is likely to see much larger impacts overall. A Ridge regression procedure to assess the correlations between site-specific covariates and treatment effects indicates that the remaining heterogeneity is strongly correlated with differences in economic variables, but not with differences in study design protocols. The average interest rate and the average loan size have the strongest correlation with the treatment effects, and both are negative. 
\end{abstract}

\newpage

\section{\large Introduction}
Researchers and policymakers increasingly have access to results from several experimental studies of the same phenomenon. The question of how to aggregate the results of multiple experiments across different contexts is now pertinent. Different studies of the same policy or intervention often produce different results, but both the extent of the true variation in the underlying treatment effects and the source of such variation are often unclear. While there is a growing understanding of the need to aggregate across studies and assess this underlying variation, and several attempts at cross-study aggregation already in the literature (e.g. Vivalt 2016, Pritchett and Sandefur 2015),  there is currently no consensus in economics regarding the appropriate methodology. But there is a statistical methodology which is ideally suited to aggregating evidence and assessing the variation in effects across study sites, and has already been used for this purpose by statisticians: Bayesian hierarchical models and their associated metrics of heterogeneity (Rubin 1981, Gelman et al 2004). In this paper I apply this methodology to the data from seven randomized controlled trials of microcredit expansions.

There are several efforts to aggregate evidence from the various microcredit studies in the form of review articles, such as Banerjee (2013) or Banerjee et al (2015a). This relatively informal approach has the advantage of incorporating expert judgment, but offers no clear way to keep track of the multiple dimensions of heterogeneity between the studies. As a result, review articles often employ simple but misleading aggregation techniques such as ``vote counting'' the statistically significant and insignificant results - for examples see Sandefur (2015) or Banerjee et al (2015a), for a critique of vote counting see Hedges and Olkin (1980) or section 9.4.11 of the Cochrane Handbook (Higgins and Green 2011). Formal aggregation methods can avoid these heuristics and keep track of the differences across studies more rigorously. 

Yet formally aggregating the evidence from studies performed in different countries with different implementation and experimental protocols is a challenging task. The interventions in the microcredit literature are fundamentally similar but not exactly the same, which is almost always the case in economics. The economic and social contexts of the study sites are different, but the extent to which these differences affect the experimental outcomes is unknown. Computing the arithmetic mean of the estimated treatment effects from each study does not capture our best understanding of the evidence, not even if the estimates are weighted inversely to their standard errors or other sample variability metrics. On the one hand, if these site-specific treatment effects are very different then averaging or ``pooling'' them is not a useful exercise, as this average does not describe a coherent population object. But on the other hand, if the effects are similar enough that we can learn something across contexts, then it is inefficient even to compute these site-specific effects in isolation from one another, and we should use all the data to adjust our estimate the effect in each site. 

In economics, researchers often have access to the actual datasets from the randomised controlled trials (RCTs) we seek to aggregate. Many economics journals, particularly journals published by the \emph{American Economics Association} such as \emph{The American Economic Journal: Applied Economics}, require the experimental data to be published alongside the studies. Most formal techniques for aggregation and meta-analysis use the reported point estimates and standard errors as their input data, because meta-analysts in statistics and other fields typically did not have access to the underlying study's ``microdata''. In principle, access to the microdata allows for a more comprehensive and detailed analysis of the evidence. It may be that heterogeneity in the observed effects of an intervention reflect contextual differences between the study protocols, national or local environments, or even the composition of types of households in each site. It is not possible to explore the role of these covariates without access to the microdata.


This paper constitutes a first attempt to use Bayesian hierarchical models to aggregate the microdata from multiple RCTs in economics. These models are well suited to the address the challenges of aggregation across heterogeneous contexts, and have been used in statistics and medicine since at least 1981 (see Rubin 1981, Gelman et al 2004). They are now being adopted into economics as a result of the increasing availability of multiple RCTs of similar interventions, but have not yet been applied to microdata (see Burke et al 2014, Vivalt 2015). In this paper I aggregate and synthesise the results from all existing RCTs of expanding access to microcredit: Angelucci et al (2015), Attanasio et al (2015), Augsberg et al (2015), Banerjee et al (2015b), Crepon et al (2015), Karlan and Zinman (2011), and Tarozzi et al (2015). Due to the policies of the two journals that published these papers - the \emph{AEJ:Applied} and \emph{Science} - all the microdata from these RCTs is freely available online. I fit Bayesian hierarchical models to the microdata from these studies to estimate the set of site-specific treatment effects on outcomes at the household level, as well as the general treatment effect common to all sites. The models are equipped with several metrics to quantify the strength of the relationship between the site-specific effects, and thus the relative importance or predictive power of the general treatment effect for the set of broadly comparable sites.

The results suggest that the effect of microcredit access is likely to be positive but small in magnitude relative to control group average levels, and the possibility of a negative impact cannot be ruled out.  I find that the site-specific effects are strongly related with 60\% pooling on average, indicating that the generalized effect is a reasonably informative object. This suggests reasonably high external validity within the class of comparable sites, although there is some remaining heterogeneity. Splitting the treatment effect apart according to contextual variables at the household level reveals that the detected heterogeneity in effects across sites is almost entirely driven by heterogeneous effects for the 27\% of households who operated a business prior to microcredit expansion. A Bayesian Ridge procedure to assess the correlation between treatment effects and study-specific contextual variables indicates that economic variables such as the average interest rate and loan size are more predictive of differences in treatment effects than study protocol variables such as the unit of randomization. These results differ substantially from the conclusions drawn in informal review articles such as Banerjee et al (2015a) and previous attempts to formally aggregate evidence on microcredit such as Pritchett and Sandefur (2015) and Vivalt (2016) which failed to separate sampling variation in the estimates from genuine underlying heterogeneity.

\section{\large Methodology}\label{section: methodology}

\subsection{Bayesian Hierarchical Models}
The Bayesian hierarchical approach to multi-study aggregation is built on the model used in Rubin (1981) to analyze the results of several parallel experiments. The model is concerned with $K$ studies or ``sites'' in which researchers performed similar interventions and measured the impact on similar outcomes. The model is fit to the set of estimated treatment effects reported in the $K$ different studies, denoted $\{\hat{\tau}_k\}_{k=1}^{K}$, and their estimated standard errors $\{\hat{se}_{\tau_k}\}_{k=1}^{K}$. The core of the model is a hierarchical structure in which each site has its own treatment effect, $\tau_k$, but these effects are all drawn from a common ``parent distribution'' governed by an unknown mean and variance parameters ($\tau, \sigma^2_{\tau}$). The Rubin (1981) model uses a Normal-Normal structure:
\begin{equation}
\begin{aligned}
\hat{\tau}_k &\sim N(\tau_k, \hat{se}_k^2) \; \forall \; k \\
\tau_k &\sim N(\tau, \sigma_{\tau}^2) \; \forall \; k .
\end{aligned}
\label{rubin model}
\end{equation}

The functional form of the lower level of the likelihood is simply the sampling distribution of the estimators used in each site, and the formulation in Rubin (1981) applies as long as the estimator is unbiased and asymptotically normal. As RCTs in economics are typically analyzed using OLS regressions, and the analysis of each RCT typically makes the assumption of unbiasedness and asymptotic Normality to calculate their standard errors, the lower level functional form imposes no more structure than the original papers. The choice of functional form for the distribution in the upper level of the likelihood is less obvious. The Normal was chosen for Rubin (1981) for tractability and because it has attractive frequentist  properties, delivering lower mean squared error for estimating the set $\{{\tau}_k\}_{k=1}^{K}$ relative to other options (Efron and Morris, 1977).

The model above can be generalized using various functional forms and can easily include other pieces of information, as long as all $K$ studies report them. For the task of modeling heterogeneity in the impact of microcredit, the value of the control group mean $\mu_k$ is plausibly related to the size of the treatment effect $\tau_k$ in each site, though the sign and magnitude of the correlation is unknown. This is often the case in economics and so I propose to incorporate this useful information, both to improve our inference on the treatment effects and to try to detect and understand the correlation here. The estimated control mean $\hat{\mu}_k$ along with its standard error $\hat{se}_{\mu_k}$ can be modeled as follows:
\begin{equation}
\begin{aligned}
\hat{\tau}_k &\sim N(\tau_k , \hat{se}_{\tau_k}^2) \; \forall \; k \\
\hat{\mu}_k &\sim N(\mu_k , \hat{se}_{\mu_k}^2) \; \forall \; k \\
\left( \begin{array}{c}
\mu_{k}\\
\tau_{k}
\end{array} \right) 
&\sim
N\left( \left(
\begin{array}{c}
\mu\\
\tau
\end{array} \right), V \right) \; \text{where} \;V = \left[ \begin{array}{cc} \sigma^2_{\mu} & \sigma_{\tau\mu} \\ \sigma_{\tau\mu} & \sigma_{\tau}^2 \end{array} \right]\forall \; k. \\
\end{aligned}
\label{generalised rubin model}
\end{equation}

Although this model can be estimated using the reported parameters, I have access to the full data from the seven studies of microcredit expansions. Hence, I can fit a hierarchical regression model directly to the study outcomes in the spirit of the Rubin (1981) model. Consider some outcome of interest, such as profits or consumption for a household $i$ in study site $k$, denoted $y_{ik}$. Denote the binary indicator of treatment status by $T_{ik}$. Allow the variance of the outcome variable $y_{ik}$ to vary across sites, so $\sigma_{y_k}^2$ may differ across $k$. Then the following full data model captures the key structure of Rubin (1981) and can be fit to the microdata from all $K$ studies:

\begin{equation}
\begin{aligned}
y_{ik} &\sim N(\mu_k + \tau_k T_{ik} , \sigma_{yk}^2)\; \forall \; i,k \\
\left( \begin{array}{c}
\mu_{k}\\
\tau_{k}
\end{array} \right) 
&\sim
N\left( \left(
\begin{array}{c}
\mu\\
\tau
\end{array} \right), V \right) \; \text{where} \;V = \left[ \begin{array}{cc} \sigma^2_{\mu} & \sigma_{\tau\mu} \\ \sigma_{\tau\mu} & \sigma_{\tau}^2 \end{array} \right]\forall \; k. \\
\end{aligned}
\label{full data model}
\end{equation}

Using the microdata, it is possible possible to further explore the heterogeneity across settings using covariates either at the household level or at the site level.\footnote{If intermediate levels are specified in the data, such as a village, district or city, there may also be important variation along these dimensions that could be incorporated. If indeed these are important then the above model will underestimate the correlations between households in these areas, and will have misleadingly small posterior intervals in that case. However not all the microcredit studies have such intermediate units.} In the microcredit studies,  researchers identified several important covariates such as a household's previous business experience (Banerjee et al 2015b, Crepon et al 2015). It would be informative to know how much of the variation across settings is due to variation in composition of the households in each sample. This exercise is possible with the microdata even if these interactions models were not reported in all of the original papers, as long as the covariates were recorded. Moreover, because the subgroup analyses from one paper can be extended to the rest of the papers, this sheds light on how general or replicable any detected subgroup effect really is. Consider $L$ relevant covariates, and denote these covariates $X_{ik}$ for household $i$ in site $k$. To specify a full interactions model - that is, to examine the power set of subgroups - we now have $2^L$ intercept terms and $2^L$ slope terms, henceforth indexed by $l$ with a slight abuse of notation. There are many possible statistical dependence structures between the various treatment effects and means across sites and subgroups that can be built on framework of equations \ref{full data model}. Below is one that is quite tractable, although it is restrictive in that it enforces independence across the treatment effects in the $2^L$ subgroup blocks. Here $X_{ik}$ are all binary, so let $\pi(l) : \{1,2,\dots, 2^L\} \rightarrow \{0,1\}^L$ be the bijection that defines the full set of interactions of these variables. For $I \in \{0,1\}^L$, denote $X^{I}_{ik} = \prod_{l=1}^L [X_{ik}^{l}]^{\mathbbm{1}\{I_{l}=1\}}$, so that the likelihood is:

\begin{equation}
\begin{aligned}
y_{ik} &\sim N\left( \sum_{l=1}^{2^L}[\mu_k^l + \tau^l_k T_{ik}]X_{ik}^{\pi(l)} , \sigma_{yk}^2\right)\; \forall \; i,k \\
\left( \begin{array}{c}
\mu^l_{k}\\
\tau^l_{k}
\end{array} \right) 
&\sim
N\left( \left(
\begin{array}{c}
\mu^l\\
\tau^l
\end{array} \right), V_l \right) \; \text{where} \;V_l = \left[ \begin{array}{cc} \sigma^2_{\mu^l} & \sigma_{\tau^l \mu^l} \\ \sigma_{\tau^l \mu^l} & \sigma_{\tau^l}^2 \end{array} \right]\forall \; l, k. \\
\end{aligned}
\label{interactions model}
\end{equation}
\vskip8pt

Similarly, important covariates may exist at the site or study level. These include economic and political variables that distinguish the study sites and partner institutions from each other, as well as study protocol variables such as the unit of randomization (village versus individual). Ideally, we want to estimate the conditional distribution of the treatment effects given these variables, as this allows policymakers to form a more precise understanding of the most likely impact that an intervention would have in their specific setting.  In some cases, however, there will be more site-level covariates than sites, which means that such conditioning will not be possible without severe overfitting. In this case, the researcher can still gain an understanding of the relative importance of these covariates by employing regularization or sparsity estimation procedures at the upper level, such as Ridge or Lasso. These can be implemented within the Bayesian framework via strong priors that push the regression coefficients of these site-level variables down to zero. Denote these covariates $X$ and suppose we have $M$ such covariates that could explain or predict the variation in the treatment effect. Then the following Bayesian hierarchical model would provide this conditioning, or conditioning plus regularization, as the case may be: 

\begin{equation}
\begin{aligned}
y_{ik} &\sim N(\mu_k + \tau_k T_{ik} , \sigma_{yk}^2)\; \forall \; i,k \\
\left( \begin{array}{c}
\mu_{k}\\
\tau_{k}
\end{array} \right) 
&\sim
N\left( \left(
\begin{array}{c}
\mu + X\beta_{\mu}\\
\tau + X\beta_{\tau}
\end{array} \right), V \right) \; \text{where} \;V = \left[ \begin{array}{cc} \sigma^2_{\mu} & \sigma_{\tau\mu} \\ \sigma_{\tau\mu} & \sigma_{\tau}^2 \end{array} \right]\forall \; k. \\
\beta_{\mu} &\sim N(0,\sigma^2_{\beta_{\mu}}) \\
\beta_{\tau} &\sim N(0,\sigma^2_{\beta_{\tau}}) 
\end{aligned}
\label{site level covariates model}
\end{equation}

Conceptually, all of these models address the basic tension in aggregation across studies by specifying heterogeneous treatment effects across sites while allowing for the existence of a common component $\tau$. The hierarchical structure is agnostic about the extent to which the common component determines the treatment effect in each site, because the parameter governing its influence, $\sigma_{\tau}^2$, is itself estimated from data. By considering any $\sigma^2_{\tau} \in [0, \infty)$, the structure nests both the ``full pooling'' case in which there is no heterogeneity across sites $(\sigma_{\tau}^2 = 0)$, and the ``no pooling'' case in which the sites have no common component $(\sigma_{\tau}^2 \rightarrow \infty)$. By allowing the data to determine the most likely value of $\sigma_{\tau}^2$, hierarchical models implement ``partial pooling'' (Gelman et al 2004).

The core challenge addressed by the hierarchical framework is the separation of sampling variation from genuine heterogeneity in treatment effects across sites. This can only be done by imposing some structure on the problem. A parametric likelihood permits us to infer the genuine heterogeneity using the relative position of the site-specific treatment effects combined with information about their differing precision due to the differing variability in the outcomes across settings. This functional form is what permits us to implement the partial pooling in this flexible structure that does not take a stand $a \; priori$ on the relative size of the sampling variation and the genuine effect heterogeneity.  Without this structure, or something like it, the analyst must choose either a no pooling model which attributes all variation across studies to genuine treatment effect heterogeneity, or a full pooling model which attributes it purely to sampling variation. 

Popular ``functional form free'' approaches to analyzing multi-study data rely on these stronger assumptions. Computing a precision-weighted average of the $K$ estimated treatment effects is a full pooling technique, as is pooling the data and running one regression via ordinary least squares. These approaches will underestimate the heterogeneity across sites if the strict full pooling assumption is false. Running $K$ different regressions via ordinary least squares is a no pooling model, and the variability in the set $\{\hat{\tau}_k\}_{k=1}^{K}$ will overestimate the heterogeneity across sites if the strict no pooling assumption is false. This overestimation occurs both because the $K$ separated regressions fail to allow inference across settings via partial pooling, and because the procedure implicitly attributes all of the variation to genuine underlying heterogeneity (for an example of such analysis, see Pritchett and Sandefur 2015). Thus, while the parametric likelihood makes the model appear more structured than the typical econometric analyses of randomized trials, in fact this set up allows us to dispense with these more restrictive structures and assumptions.

These models do require that the treatment effects be ``exchangeable'' in order to perform well, which means that their joint distribution must be invariant to permutation of the $K$ indices. For example, this means that we do not have any knowledge of the ordering of the treatment effects - although we know that they may be different, and we can think of reasons why one site may have a larger or smaller effect than another site,  we do not actually know how the sites will be ordered until we see the data. If we know that a covariate should be correlated with the treatment effects, we can use the model in equation \ref{site level covariates model} which will then require $conditional$ exchangeability. Overall this requirement means that we will only be able to assess external validity, and generalizability, for the set of sites which are in fact exchangeable. Any future site for which we wish to predict or infer the treatment effect must be exchangeable with the set of sites we have already studied. Using only the RCT data means that we can generalize to the set of study sites that could plausibly have been the location of an RCT of the intervention we study, but we cannot transport our results outside of this set. Confining ourselves to this interpretation allows us to avoid the site-selection issues that arise when attempting to extrapolate to sites that are too fundamentally different to the sites studied in the literature.

\subsection{Estimation}\label{subsection:estimation}

Estimating the unknown parameters specified in the hierarchical likelihoods of models 2.1-2.4 is challenging because the likely values of the parameters on the lower level are influenced by the values of the parameters at the upper level, which introduces ripples in the likelihood. In theory either Maximum Likelihood methods or Bayesian methods can be used, but in practice there are strong reasons to prefer Bayesian inference for this problem. The primary issue with Maximum Likelihood is that to get tractability the estimation is done via ``Empirical Bayes'', which first estimates the upper level parameters and then plugs these point estimates into the lower level to estimate the lower level parameters. By conditioning on a single value of the hyperparameters, this procedure systematically underestimates the uncertainty at the lower level of the model. By contrast, Bayesian inference proceeds via estimation of the full joint posterior distribution of all unknown parameters simultaneously, from which the marginal distributions provide accurate uncertainty intervals. 

The Bayesian approach does not require the compromises made by the MLE method for tractability because it performs estimation using a powerful simulation technique called Markov Chain Monte Carlo methods. These methods require a proper posterior distribution as the target distribution, which typically necessitates the use of proper prior distributions on the unknown parameters. These priors also allow the researcher to improve the estimation by targeting regions of the parameter space that are more likely to contain relevant values; if only vague knowledge of this is obtainable, then the priors can be made quite diffuse (sometimes called ``weakly informative''). If substantial expert knowledge of the likely values is available before seeing the data, this can of course be incorporated via stronger priors. Even if the prior distributions are incorrectly centered, sufficiently diffuse priors can still improve the mean squared error of the estimation by reducing the variance at the cost of some increase in bias - that is, the prior regularizes the estimates. 

Bayesian inference also provides a framework for decision-making about policy and future research that has no counterpart in frequentist inference. Indeed, our goal itself is underpinned by Bayesian thinking: we seek to update our understanding of the unknown parameters in one location using the information about the parameters from other locations. Moreover, if we wish to make decisions accounting for our uncertainty about unknown parameters, the correct object to take expectations over is the posterior distribution of the unknown, not the sampling distribution of an estimator of the unknown. Because the object of interest for policymakers is the distribution of the treatment effect in a hypothetical future site, $\tau_{K+1}$, this distribution must be computed accounting for the full joint posterior uncertainty rather than conditioning on a particular point estimate or even a particular interval estimate - the Bayesian approach provides this in the form of posterior predictive inference, which has no frequentist equivalent.

In this paper, I perform Bayesian inference with the following priors for the main specification of the model described in equations \ref{full data model}:
\begin{equation}
\begin{aligned}
\left( \begin{array}{c}
\mu\\
\tau
\end{array} \right) 
&\sim
N\left( \left(
\begin{array}{c}
0\\
0
\end{array} \right), \left[ \begin{array}{cc} 1000^2 & 0 \\ 0 & 1000^2 \end{array} \right]  \right)\\
\sigma_{yk} &\sim U[0,100000]  \; \forall \; k\\ 
V &= diag(\theta) \Omega diag(\theta) \\
\theta &\sim Cauchy(0, 10) \\
\Omega &\sim LKJcorr(3).
\end{aligned}
\label{priors}
\end{equation}
 The decomposition of the $V$ matrix into a correlation matrix $\Omega$ and scaling factor $\theta$ follows the advice in Gelman and Hill (2007). The Cauchy(0, 10) on $\theta$ permits the scaling to vary widely, and the LKJcorr(3) in $\Omega$ is a prior over the space of all correlation matrices which favors the region around independent or uncorrelated variables (Stan Development Team, 2014). In this case, the prior is informed by economic theory that suggests this correlation could take either sign: perhaps microcredit only works for relatively prosperous entrepreneurs, but perhaps it has diminishing marginal returns. With only 7 studies, we should not update our beliefs about this correlation too dramatically, so a stronger prior towards zero is warranted. In any case, the estimation of both the hypervariance $\theta$ and the correlations in $\Omega$ is a challenging exercise and with so little data the priors will typically have an influence on the posterior inference. Indeed, it is desirable to inject prior information here rather than rely on a ``raw'' correlation computed in a low data environment, which is inherently noisy. However, as a robustness check, in this paper I will also fit the classical Rubin (1981) model, whose inferences are not sensitive to the priors when they are sufficiently diffuse. I will also fit a version of the main specification (equations \ref{full data model}) which  imposes independence between the sets $\{\tau_k\}_{k=1}^K$ and  $\{\mu_k\}_{k=1}^K$, thereby eliminating that sensitivity. While sensitivity to priors can complicate the inference, the elimination of the sensitivity can only be done using restricted functional forms, so it is best to examine the full set of results rather than relying on the simpler models.

The posterior distribution for the basic full-data model is proportional to the product of the likelihood in equations \ref{full data model} and the prior in equations \ref{priors}:
 \begin{equation}
\begin{aligned}
p(\tau, \mu, \tau_1, \tau_2, \dots| Y)\propto \;& \Pi_{i =1}^{N} \Pi_{k=1}^{K} (N(y_{ik}|\mu_{k} + \tau_{k}T_{ik}, \sigma^{2}_{yk}))\\
&\times \Pi_{k=1}^{K} (N((\mu_k , \tau_k) | (\mu, \tau), V) \\
&\times N((\mu, \tau) | (0,0), I_{2}) \times Cauchy(0, 2.5) \times LKJcorr(2)
\end{aligned}
\label{posterior}
\end{equation}
This is not a known distribution, but it can be fully characterized via simulation using Markov Chain Monte Carlo methods (MCMC). The basic intuition behind MCMC methods is the construction of a Markov chain which has the posterior distribution as its invariant distribution, so that in the limit, the draws from the chain are ergodic draws from the posterior. This chain is constructed by drawing from known distributions at each ``step'' and using a probabilistic accept/reject rule for the draw based on the posterior distribution's value at the draw. 

I use a particular subset of MCMC methods called Hamiltonian Monte Carlo (HMC) methods throughout this paper, which are particularly suited to estimating hierarchical models (Betancourt and Girolami, 2013).  HMC uses discretized Hamiltonian dynamics to sample from the posterior, and has shown good performance especially combined with the No-U-Turn sampling method (NUTS) to auto-tune the step sizes in the chain (Gelman and Hoffman, 2011). HMC with NUTS is easy to implement because it can be done automatically in Stan, which is a free software module that calls C++ to fit Bayesian models from R or Python (Stan Development Team, 2014). Stan often requires no more input from the user than typing the equations for the likelihood and priors, although more complex models benefit from code written more efficiently than that. Stan automatically reports the posterior means (eg. $\tilde{\tau}$ for $\tau$) and their marginalized posterior variances (eg. $\tilde{se}_{\tau}^2$), supplying both the parameter values most likely to  be true given the data and the degree of certainty we should have about their value. Stan also automatically reports the marginal 95\% credible intervals and 50\% credible intervals.

Stan also computes and reports several performance metrics and convergence diagnostics for the HMC in every model it fits. First, it reports the Monte Carlo error of the posterior mean, which should be small relative to the magnitude of the mean if the sampler has converged. Second, it computes the $\hat{R}$ metric of Gelman and Rubin (1992) by randomly perturbing the starting points for the HMC chains and then checking the between variance of the chains relative to the within-chain variance. If all the chains have converged to the posterior, their within variance should be the same as their between variance: the $\hat{R}$ is the ratio of these variances and should be close to 1. For each model, I run 4 chains and accept $\hat{R} < 1.1$. 

\subsection{Pooling Metrics}

Bayesian hierarchical models come equipped with several natural metrics to assess the extent of pooling across sites shown in the posterior distribution, developed and studied by statisticians (Gelman et al 2004, Gelman and Pardoe 2006).  In the context of multi-study aggregation, the extent of pooling across study sites has a natural interpretation as a measure of external validity. The extreme case of full pooling $(\sigma^2_{\tau} = 0)$ corresponds to perfect external validity wherein all $\tau_k = \tau$, so by conducting a study in one site we learn as much about the treatment effect for all $K$ sites as we do for the specific site we study. The estimate may be noisy or have other problems, but it is equally valid for site $k$ as for site $k'$.  The no pooling case, where $\tau$ is an uninformative object $(\sigma^2_{\tau} \rightarrow \infty)$, corresponds to zero external validity because we learn nothing about site $k'$ from site $k$. An obvious metric of external validity in this framework is therefore the magnitude of $\sigma^2_{\tau}$, and a good estimate for it is the posterior mean denoted $\tilde{\sigma}^2_{\tau}$. 

The drawback of using $\tilde{\sigma}^2_{\tau}$ as a pooling metric is that it is not clear what exactly constitutes a large or small value of this parameter in any given context. Thus, while it is important to report and interpret $\tilde{\sigma}^2_{\tau}$ , it is also useful to examine pooling metrics whose magnitude is easily interpretable. These include the conventional ``pooling factor'' metric, defined as follows (Gelman and Hill 2007, p. 477):
\begin{equation}
\begin{aligned}
\omega(\tau_k) = \frac{ \hat{se}_k^2}{\tilde{\sigma}_{\tau}^2 + \hat{se}_k^2}.
\end{aligned}
\label{pooling factor}
\end{equation}
This metric has support on [0,1] because it decomposes the potential variation in the estimate in site $k$ into genuine underlying uncertainty and sampling error. It compares the magnitude of $\tilde{\sigma}^2_{\tau}$  to the magnitude of $\hat{se}_k^2$, the sampling variation in the separated estimate of the treatment effect from site $k$. Here, $\omega(\tau_k) > 0.5$ indicates that $\tilde{\sigma}^2_{\tau}$ is smaller than the sampling variation, indicating substantial pooling of information and a ``small'' $\tilde{\sigma}^2_{\tau}$. If the average of these $K$ pooling metrics across sites is above 0.5, then this suggests the genuine underlying heterogeneity is smaller than the average sampling variance. In that case, $\tau_k$ is a better signal of $\tau$ than $\hat{\tau}_k$ is of $\tau_k$, and if we are comfortable using our no-pooling model for each site, we should be comfortable extrapolating to the general case. 

The fact that the $\omega(\tau_k)$ uses sampling variation as a comparison is both a feature and a drawback. In one sense this is exactly the right comparison, since we are scoring how much we learned about site $k'$ by analysing data from site $k$ against how much we learned about site $k$ by analyzing data from site $k$, which is captured by the sampling variation in $\hat{\tau}_k$. Yet in another sense, if the sampling variation is very large or small due to an unusually small or large sample size or level of volatility or noise in the data, it may be beneficial to have an alternative pooling metric available. There are two additional metrics I consider in this paper which make for useful alternatives. The first such metric is a ``brute force'' version of the conventional pooling metric, which I define as follows:
\begin{equation}
\begin{aligned}
{\omega_b}(\tau_k) \equiv \{ \omega: \; \tilde{\tau}_k = {\omega} \tilde{\tau} + (1-{\omega})\hat{\tau}_k \}.
\end{aligned}
\label{brute force pooling factor}
\end{equation}
This metric scores how closely aligned the posterior mean of the treatment effect in site $k$, denoted $\tilde{\tau}_k$, is to the posterior mean of the general effect $\tilde{\tau}$ versus the separated no-pooling estimate $\hat{\tau}_k$. Here,  $\omega_{b}(\tau_k) > 0.5$ indicates that the generalized treatment effect is actually more informative about the effect in site $k$ than the separated estimate from site $k$ is for site $k$ (because $\tilde{\tau}_k$ as our best estimate of the effect in site $k$). The motivation for computing this $\omega_{b}(\tau_k)$ is that in the Rubin (1981) model it is actually identical to the conventional pooling metric, but it is not identical in more complex models that pool across multiple parameters (such as model \ref{full data model}). Solving for ${\omega_b}(\tau_k)$ by simple algebra is a ``brute force'' approach which provides a useful additional metric in these more complex models. I manually constrain it to take values between $[0,1]$ as the rare occasions on which it falls outside this range are due to shrinkage on other parameters rather than due to any feature of the parameters in question.

Another pooling metric that can be computed for these models is the ``generalized pooling factor'' defined in Gelman and Pardoe (2006), which takes a different approach using posterior variation in the deviations of each $\tau_k$ from $\tau$. Let $E_{post}[.]$ denote the expectation taken with respect to the full posterior distribution, and define $\epsilon_k = \tau_k - \tau$. Then the generalized pooling factor for $\tau$ is defined:
\begin{equation}
\begin{aligned}
\lambda_{\tau} \equiv 1-\frac{\frac{1}{K-1}\sum_{k=1}^{K}( E_{post}[\epsilon_k] - \overline{E_{post}[\epsilon_k]})^2}{E_{post}[\frac{1}{K-1}\sum_{k=1}^{K}(\epsilon_k - \bar{\epsilon_k})^2]}.
\end{aligned}
\label{gelman pardoe}
\end{equation}
The denominator is the posterior average variance of the errors, and the numerator is the variance of the posterior average error across sites. If the numerator is relatively large then there is very little pooling in the sense that the variance in the errors is largely determined by variance across the blocks of site-specific errors; if the numerator is relatively small then there is substantial pooling. Gelman and Pardoe (2006) suggest interpreting $\lambda_{\tau} > 0.5$ as indicating a higher degree of general or ``population-level'' information relative to the degree of site-specific information. 


For policy purposes, the most relevant metric is the total uncertainty about the treatment effect in future sites. This is captured by the distribution of the treatment effect in the next site, $\tau_{K+1}$. While economists often make conditional predictions about such objects, the Bayesian approach allows us to estimate the entire marginal posterior predictive distribution, thereby accurately characterizing the uncertainty. While the posterior inference on $\tau$ provides us with some understanding of the likely impact in any exchangeable site not yet studied, as $E[\tau_{K+1}] = \tau$, it does not provide the whole story. To understand the predictive distribution of $\tau_{K+1}$ we need to know not only $\tau$ and its posterior uncertainty but the posterior estimate of the average dispersion of the $\tau_k$ draws around this $\tau$. It is however still not enough to know the posterior mean value of $\sigma_{\tau}^2$ - we need to account for our uncertainty about this parameter as well. The Bayesian approach allows us to estimate the posterior distribution of the new parameter $\tau_{K+1}$ marginalizing over the joint posterior distribution of $\tau$ and $\sigma^2_{\tau}$: this is called the marginal posterior predictive distribution of $\tau_{K+1}$ and is the only way to accurately capture the total uncertainty about this effect.

\section{\large The General Impact of Microcredit Expansions} 


Aggregation across multiple contexts allows the estimation of the general impact of expanding access to microcredit services on household outcomes. In all the Bayesian hierarchical models from section \ref{section: methodology} this general effect is captured by $\tau$, the mean of the parent distribution from which all site-specific treatment effects are drawn. This is the expected value of the treatment effect in all sites and in any future site which is broadly comparable to the current set of sites. This is precisely the quantity of interest for policymakers who must make decisions about interventions in places that have not yet been studied in an RCT. In this section I report results for household business profit, revenues and expenditures, as well as household consumption, consumer durables spending and ``temptation goods'' spending. The latter 3 variables were unfortunately collected only by a subset of studies but are important enough in the theoretical literature on microcredit that they should be aggregated nonetheless (Banerjee 2013). 

\begin{figure}[h!]
  \centering
    \includegraphics[scale=0.6]{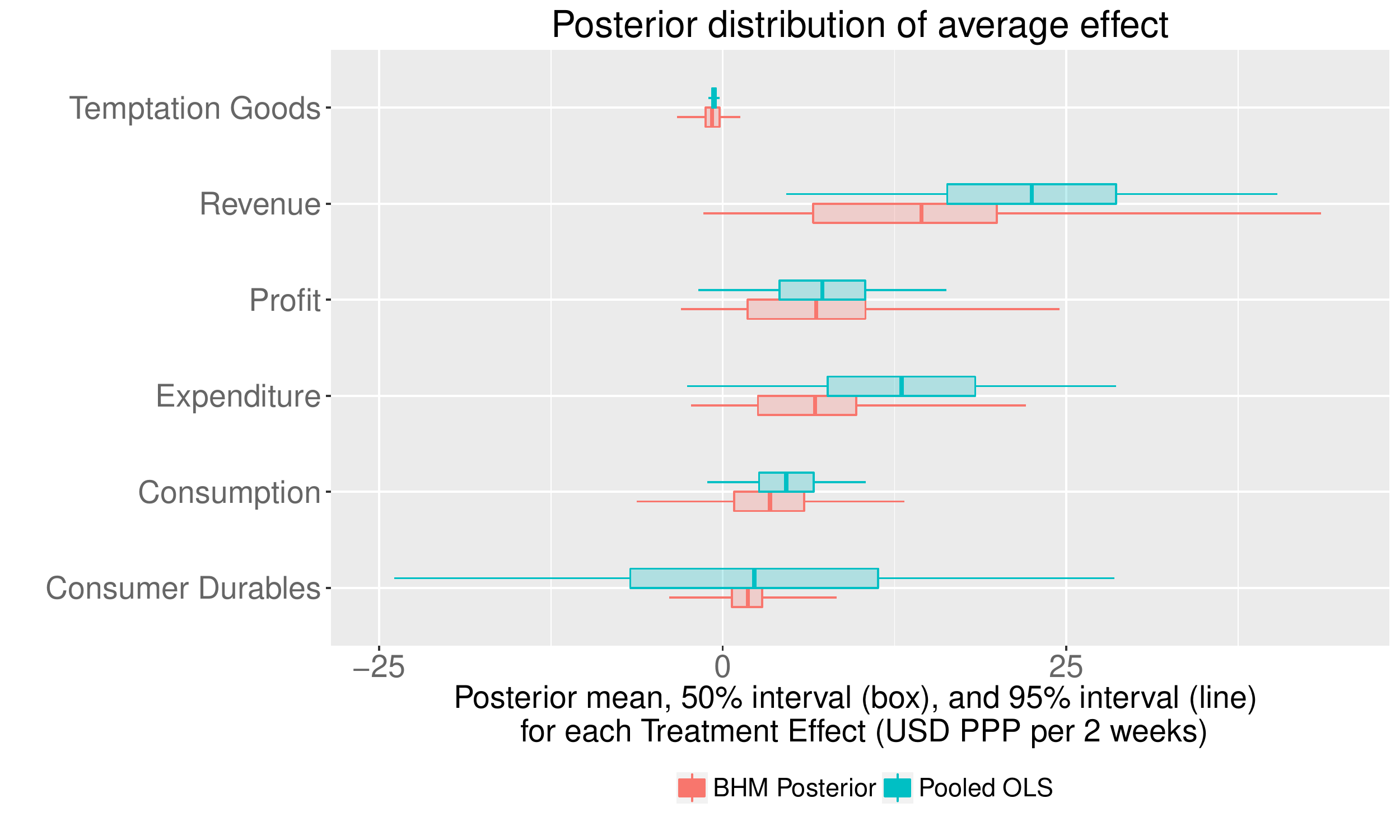}
  \caption{Graph of posteriors for each $\tau$ from the main specification of the joint model, with the full pooling OLS intervals for comparison.} \label{tau posteriors graph}
\end{figure}

\begin{figure}[h!]
  \centering
    \includegraphics[scale=0.6]{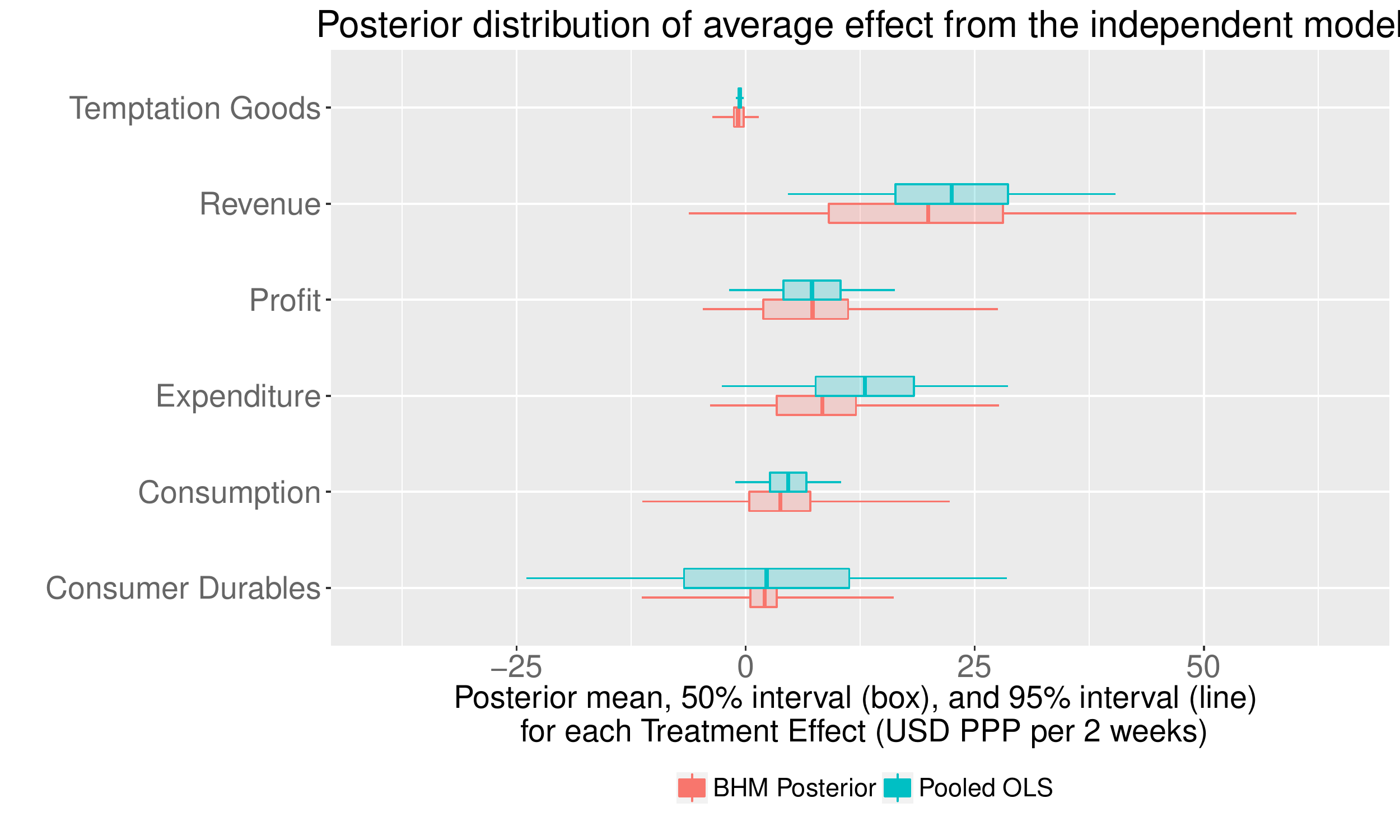}
  \caption{Graph of posteriors for each $\tau$ from the independent model, with the full pooling OLS intervals for comparison.} \label{independent tau posteriors graph}
\end{figure}

To estimate $\tau$ I fit the model described by equations \ref{full data model} in Stan to each outcome variable after standardizing all units to USD PPP over a two week period (indexed to 2010 dollars). The graph in figure \ref{tau posteriors graph} shows the posterior distributions of $\tau$ for each of the six outcomes, and for comparison, the sampling distribution of the OLS estimator for the full pooling model's estimate of $\tau$. As a robustness check I also fit the independent model specification, with results shown in figure \ref{independent tau posteriors graph}. In both cases the Bayesian hierarchical estimates find more posterior mass around zero than the full pooling model does, and tends to have wider uncertainty intervals reflecting that the model typically detects some heterogeneity across studies. These results show that the effect of microcredit is likely to be positive but small in magnitude relative to control group average levels, and the possibility of a negative impact cannot be ruled out. For example, the posterior mean $\tilde{\tau}$ for profit is about 7 USD PPP per two weeks, while the control group mean is about 95 USD PPP per two weeks and the control group standard deviation is 160 USD PPP per two weeks. In summary, while we have much more evidence of positive impact than of negative impact, the general impact is uncertain and likely to be small.

In two cases, for revenues and temptation goods, the full pooling model gives a substantially different result to the Bayesian hierarchical models. In fact, a frequentist assessment of the full-pooling OLS analysis would declare these two variables ``statistically significant'', but the hierarchical model finds that their central 95\% posterior intervals actually include zero quite comfortably. This difference arises because the full pooling model can neither detect heterogeneity nor incorporate this heterogeneity across sites into its estimate of the uncertainty about $\tau$. Even for the variables where the full pooling model would not declare ``statistical significance'', it can overestimates the magnitude of the effect, as it does for household business expenditure. However, for many variables the full pooling model and the Bayesian hierarchical model produce very similar estimates and intervals. The fact that the full pooling model produces reasonably similar results to the microcredit model for 4 of the 6 outcomes suggests quite substantial pooling and high generalizability. 


Thus, even when the full pooling model agrees with the Bayesian hierarchical by the standards of ``statistical significance'', the Bayesian model results are interpretable in a way that the full pooling model results are not. The full pooling model reports an average without any assessment of the similarity of the objects that comprise the average. The Bayesian hierarchical model performs this assessment and uses that answer to inform its estimate of the uncertainty interval - hence, the tightness of the Bayesian interval on the $\tau$ estimate already tells us something about the generalizability. If the site estimates are highly heterogeneous, this will translate into higher posterior uncertainty about the value of $\tau$ in the hierarchical model (but not in the full pooling model). If the estimates are close together then they probably lie quite close to the mean of the distribution from which they are drawn. For the microcredit data, the fact that the posterior intervals of $\tau$ tend to be somewhat wider than the full-pooling model intervals suggest that the model detects some heterogeneity among the effects - but as they are not that much wider, this heterogeneity is not substantial. The posterior intervals of $\tau$ for consumer durables and expenditure are actually tighter than the full-pooling model, reflecting strong pooling: much of the apparent heterogeneity here was due to sampling variation within sites, which the hierarchical model can separate out when estimating $\tau$. 

\begin{figure}[h!]
  \centering
    \includegraphics[scale=0.7]{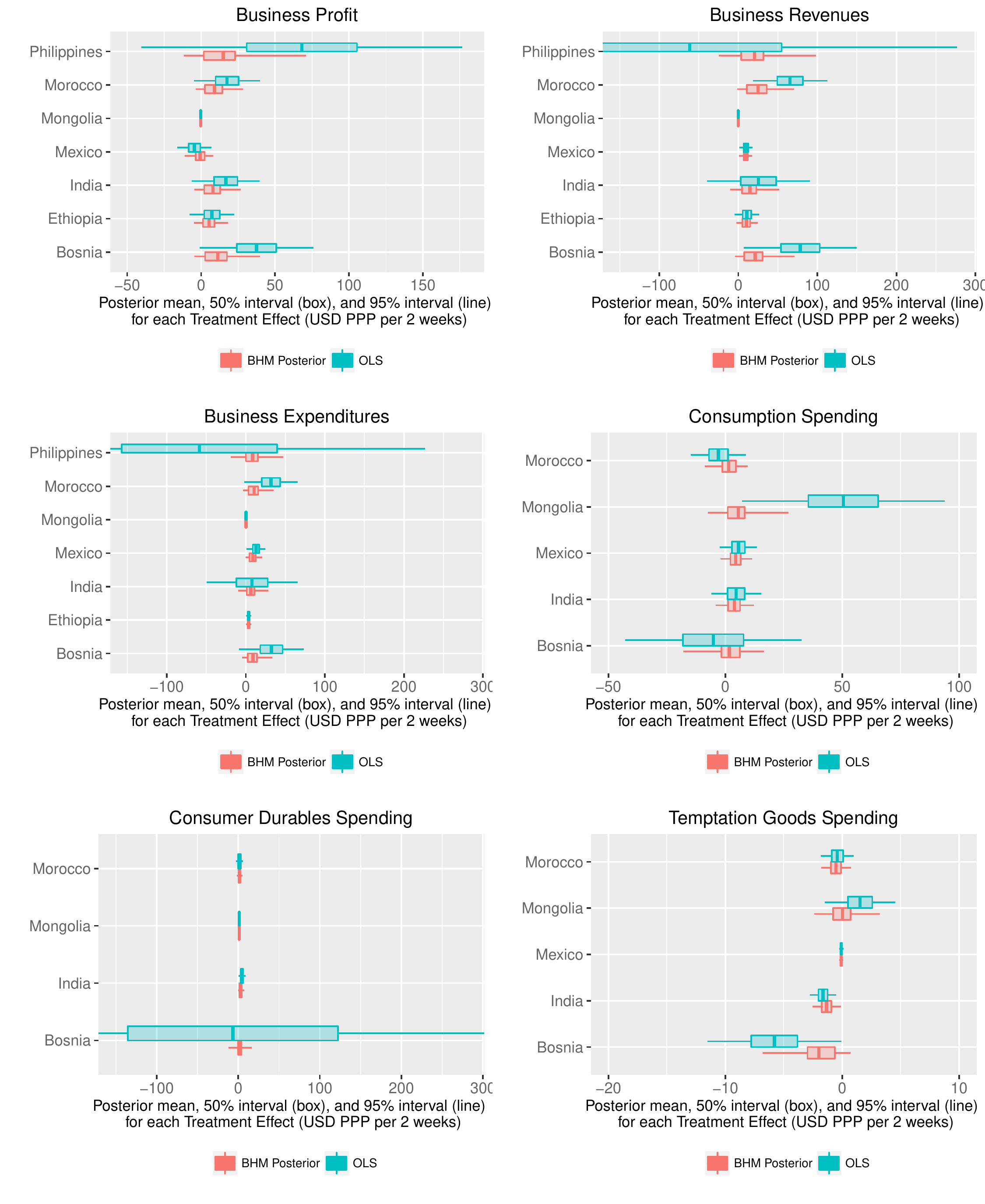}
  \caption{Graph of posteriors for each $\tau_k$ from the main specification of the joint model, with the no-pooling OLS intervals for comparison. As the scales of the sampling error differ across sites, some intervals have not been fully shown here. The tables in Appendix A provide the values of these 4 quantiles and the mean for all marginal posteriors for the main specification model.} \label{tau k posteriors with ols graph}
\end{figure}
\begin{figure}[h!]
  \centering
    \includegraphics[scale=0.7]{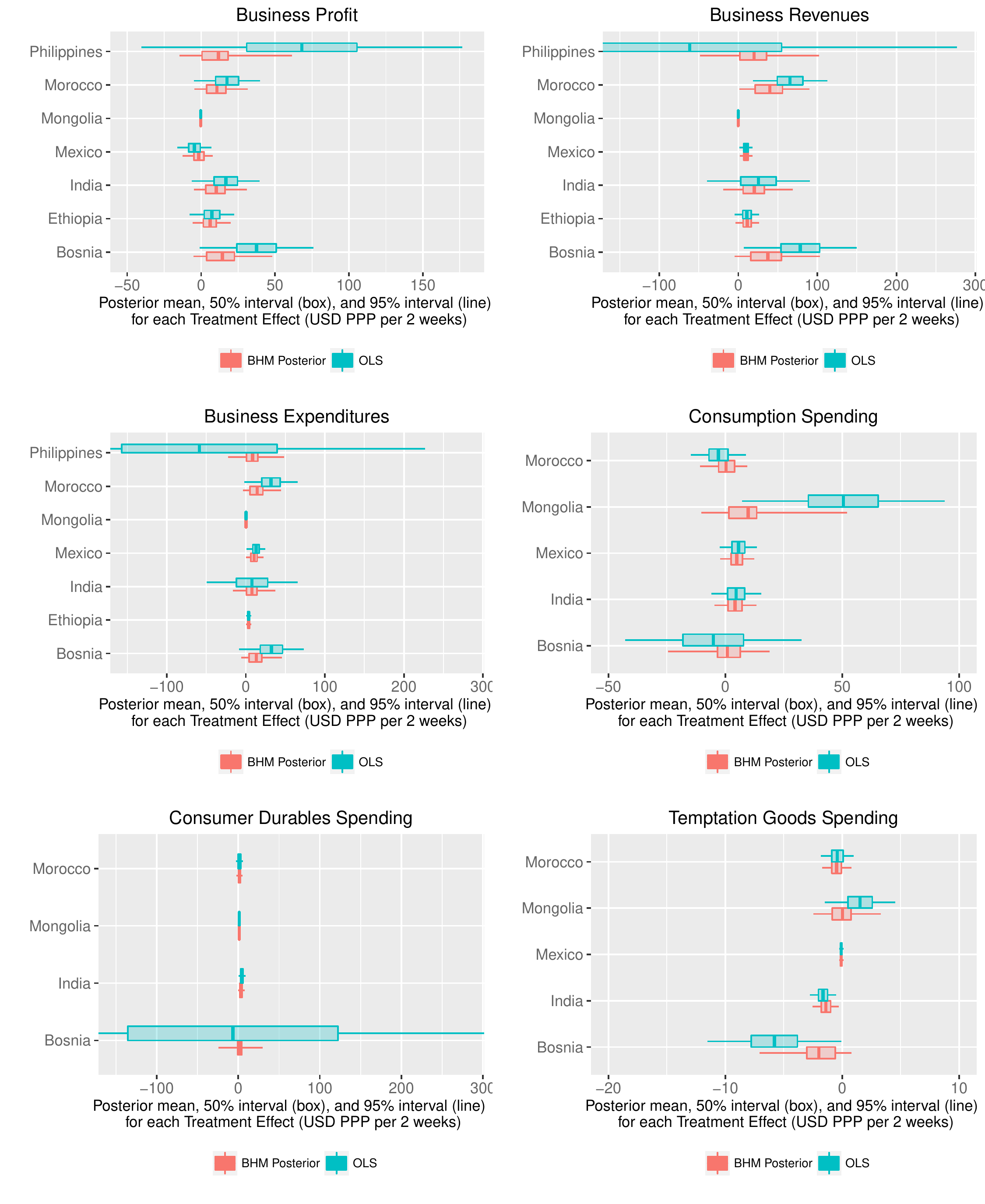}
  \caption{Graph of posteriors for each $\tau_k$ from the independent model, with the no-pooling OLS intervals for comparison.} \label{independent tau k posteriors with ols graph}
\end{figure}

To understand why the Bayesian hierarchical model consistently places more probability mass near zero than the full pooling model does for the microcredit data, it is useful to examine the study-specific treatment effects $\{\tau_k\}_{k=1}^{K}$ and their no-pooling estimates  $\{\hat{\tau}_k\}_{k=1}^{K}$ as shown in figure \ref{tau k posteriors with ols graph}. Due to the occasionally varying scales of the sampling error, not all intervals have been fully displayed graphically from end to end, but this information can be found in the tables in Appendix A. The independent model results for the same variables are shown in \ref{independent tau k posteriors with ols graph} and are very similar. In almost all cases, the more precisely estimated effects are also smaller in magnitude and typically very close to zero. It is not only that the most precise study is the smallest: the ordering of study effect sizes from smallest to largest is typically the same as the ordering of standard errors from smallest to largest. Figure \ref{tau k posteriors with ols graph} shows that there is substantial pooling for all outcomes, and the cluster of precise studies near zero pulls the less precise studies dispersed widely around them in towards zero. Temptation goods exhibits the least amount of shrinkage, reflecting its relatively low sampling variation compared to the dispersion in site-specific effects, but extreme results are still pulled somewhat towards zero. 

 These results differ from the informal analysis of Banerjee et al (2015a), which predicted that combining the six 2015 studies and running pooled regressions might find a positive and significant impact on profit. This was a reasonable hope, as four of the isolated $\hat{\tau}_k$ estimates were positive and reasonably large but statistically insignificant, and pooling does increase power and precision. But for profit, expenditures, consumption and consumer durables spending the estimate of $\tau$ from the pooled OLS and the marginal posterior from the Bayesian hierarchical model both have high density around zero. For revenues and temptation goods the full pooling model does indeed exclude zero in the 95\% interval, but the partial pooling model detects enough heterogeneity to overturn this result. This conclusion also differs from the result in Vivalt (2016), which reports a small negative treatment effect of microcredit on profit. However, that analysis aggregates a different set of microcredit studies including observational studies (and does not incorporate the control means). As we cannot be completely confident in the exclusion restrictions from the observational studies, the results of aggregating the 7 RCTs may be more reliable.

\section{\large Quantifying Heterogeneity in the Impact of Microcredit Expansions}

\begin{landscape}
\begin{figure}[h!]
  \centering
    \includegraphics[scale=0.8]{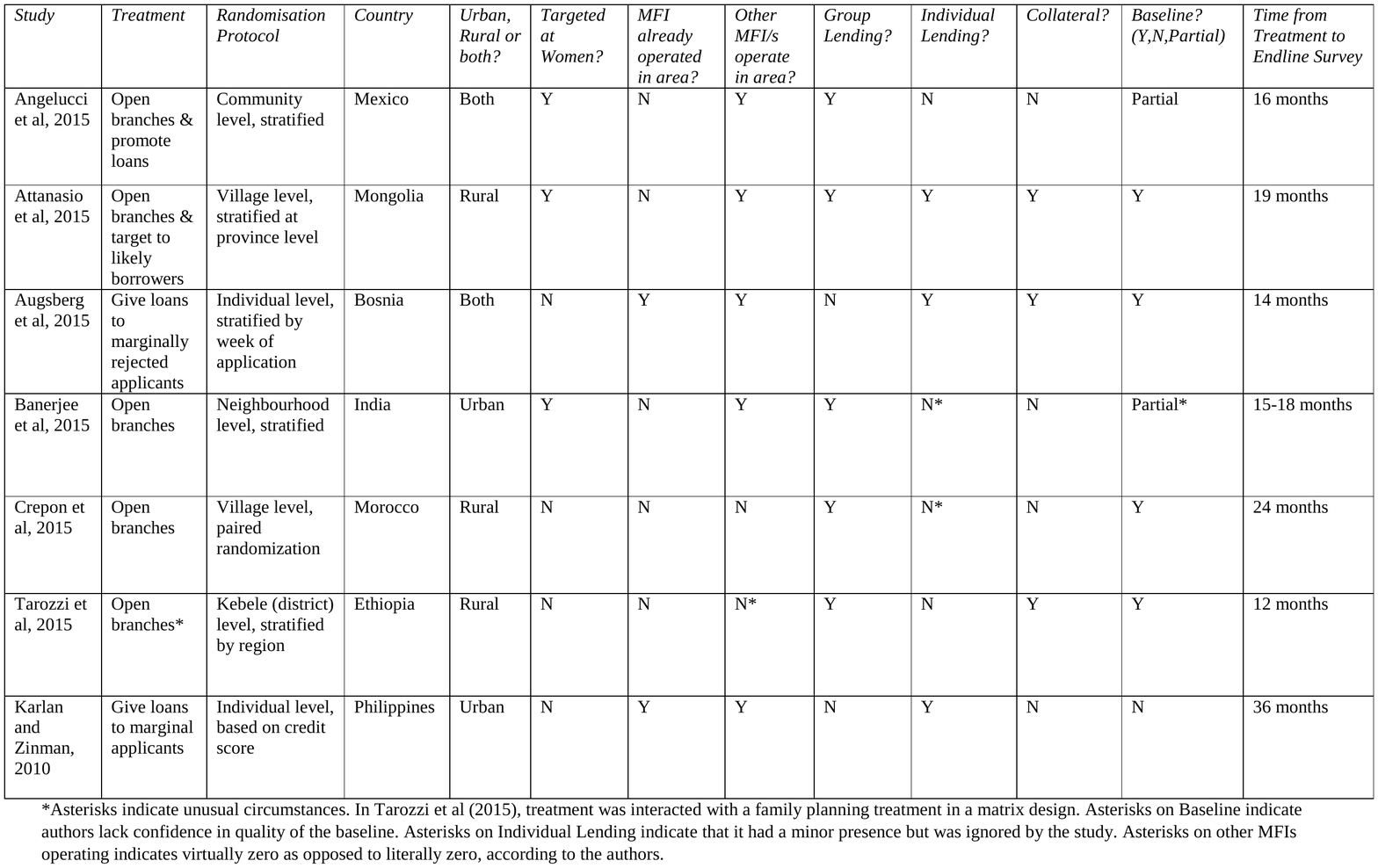}
  \caption{Summary of the 7 studies considered in this paper.} \label{summary of RCTs}
\end{figure}
\end{landscape}

The posterior inference on $\tau$ provides us with some understanding of the likely impact in a future site not yet studied, as $E[\tau_{K+1}] = \tau$, but it does not provide the whole story. To understand the predictive distribution of $\tau_{K+1}$ we need to know not only $\tau$ and its posterior uncertainty but the posterior estimate of the average dispersion of the $\tau_k$ draws around this $\tau$. This is the external validity question that motivated our use of the Bayesian hierarchical models, and for good reason: the seven microcredit studies differed in their economic contexts, study protocols, population compositions, and along variety of other dimensions. These differences are summarized in figure \ref{summary of RCTs}. Clearly, the estimated site-specific treatment effects might be somewhat heterogeneous; the key question is how heterogeneous the underlying effects really are, and how informative they are for one another. The fact that the posterior intervals of $\tau$ in the Bayesian hierarchical model are often wider than the full pooling model suggest some heterogeneity, but as they are not very much wider, the extent of the variation may be quite small. To quantify the heterogeneity in the site-specific treatment effects of microcredit expansions, I now report the metrics discussed in section \ref{section: methodology}.

\begin{figure}[h!]
  \centering
    \includegraphics[scale=1]{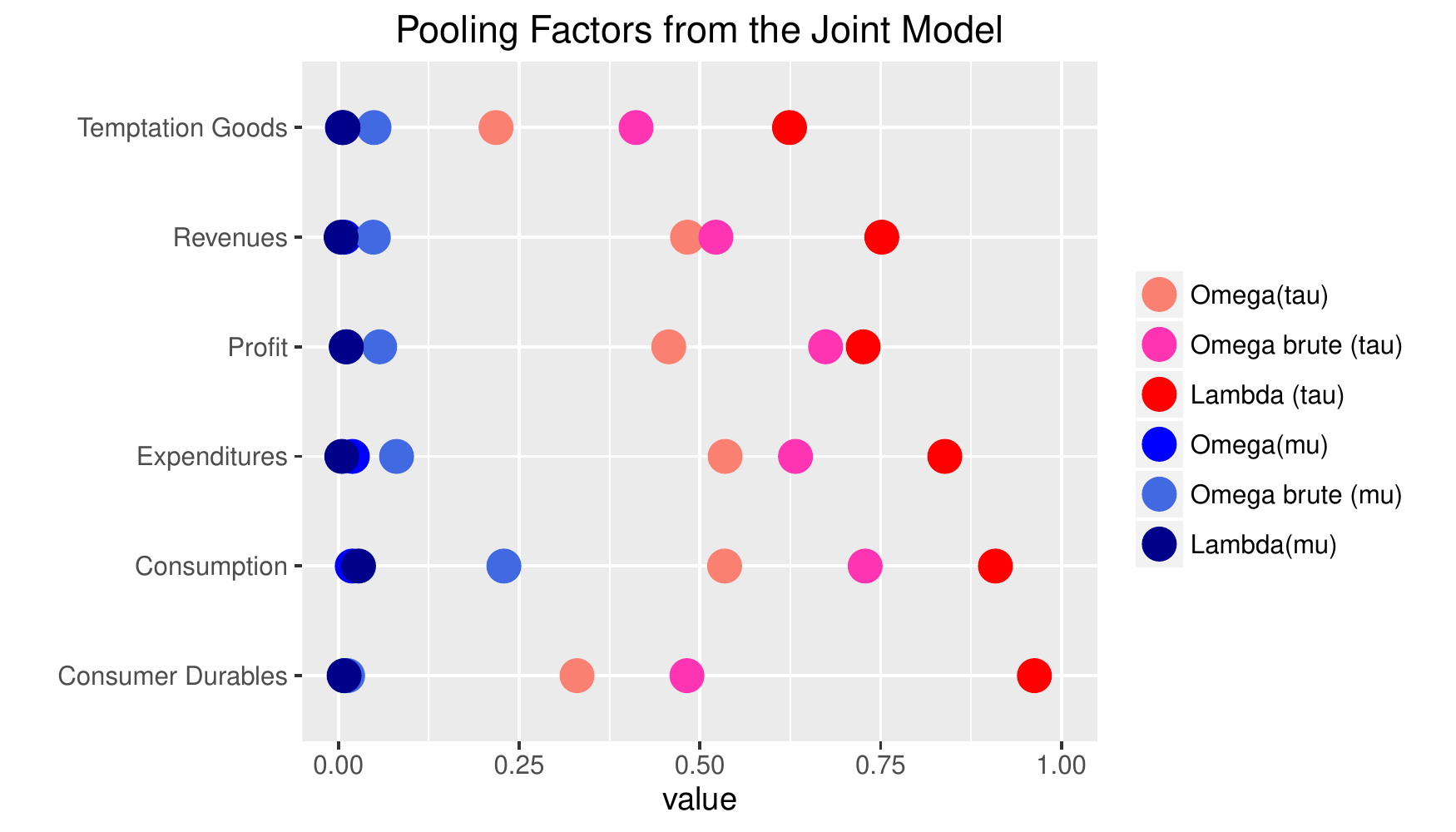}
  \caption{Pooling metrics for all outcomes for the main specification} \label{pooling metrics graph}
\end{figure}

\begin{figure}[h!]
  \centering
    \includegraphics[scale=1]{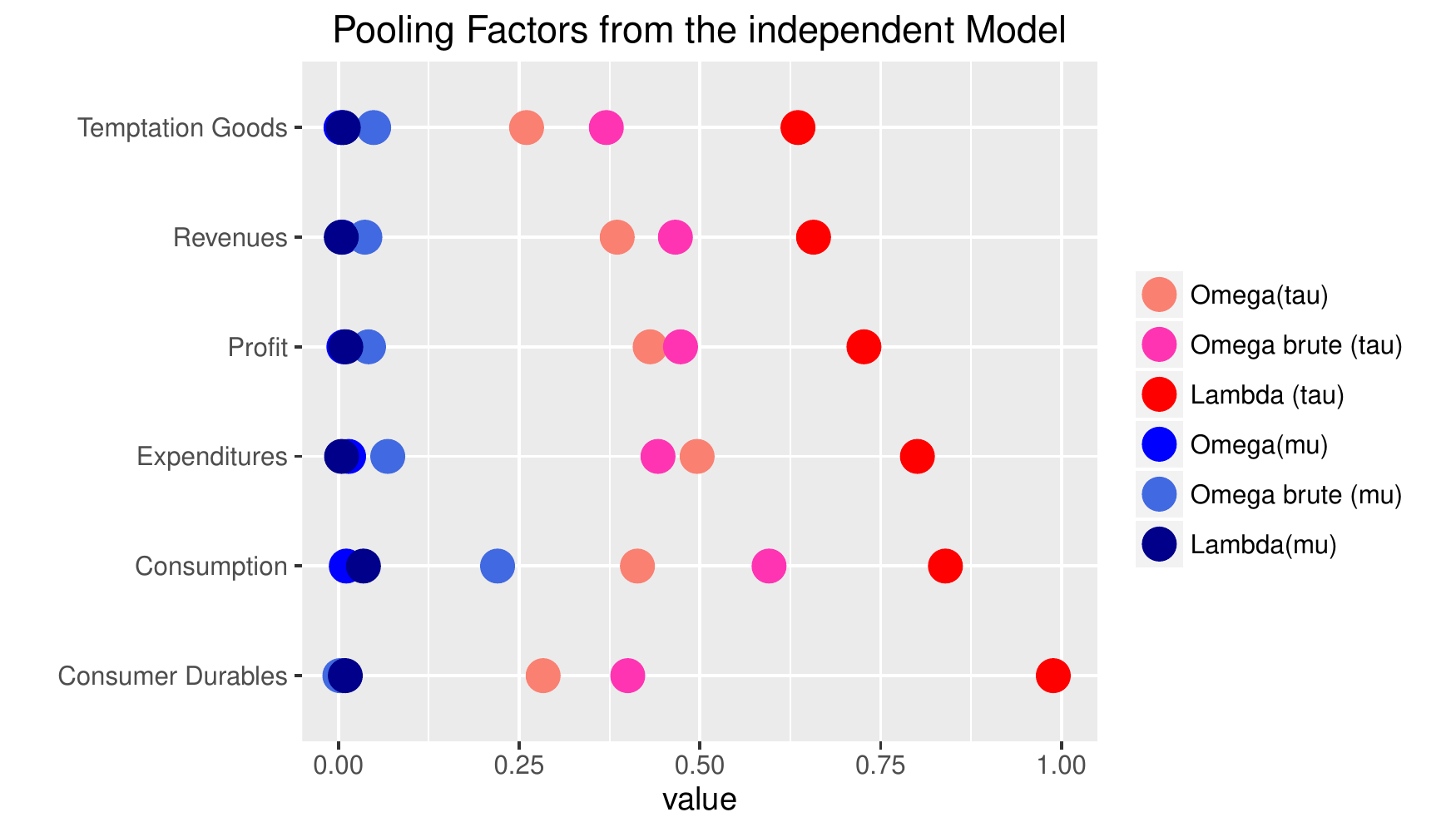}
  \caption{Pooling metrics for all outcomes for the independent model specification} \label{independent pooling metrics graph}
\end{figure}

The graphic in figure \ref{pooling metrics graph} displays the pooling metrics, $\{\omega, \tilde{\omega}, \lambda \}$ computed for both the treatment effects $\{\tau_k\}_{k=1}^{K}$  and the control group means $\{\mu_k\}_{k=1}^{K}$. As a robustness check I compute these metrics for the model that enforces independence between $\mu$ and $\tau$, and the reuslts are shown in figure \ref{independent pooling metrics graph}. In both cases I find substantial though not full pooling on the treatment effects, with an average value around 0.6 across all metrics and all outcomes. There is some heterogeneity in the metrics, and across the outcomes it seems there is more pooling on business variables than on consumption variables, but overall the pooling is significant. This indicates that $\tau_k$ tends to be a better signal of $\tau$ than the no-pooling estimate $\hat{\tau}_k$ is of $\tau_k$ in these studies. Hence, if we are comfortable using $\hat{\tau}_k$ to infer $\tau_k$ we can comfortably use any $\tau_k$ to infer $\tau$ in this literature. By contrast, I find virtually zero pooling on the control group means, with an average value of 0.03 and in many cases exactly zero. These results suggest that whatever similarities are evident in the treatment effects are not produced by pre-existing similarities in the groups before they are treated. This literature does study heterogeneous groups, yet we find reasonably similar treatment effects despite these underlying differences. 

\begin{figure}[h!]
  \centering
    \includegraphics[scale=0.6]{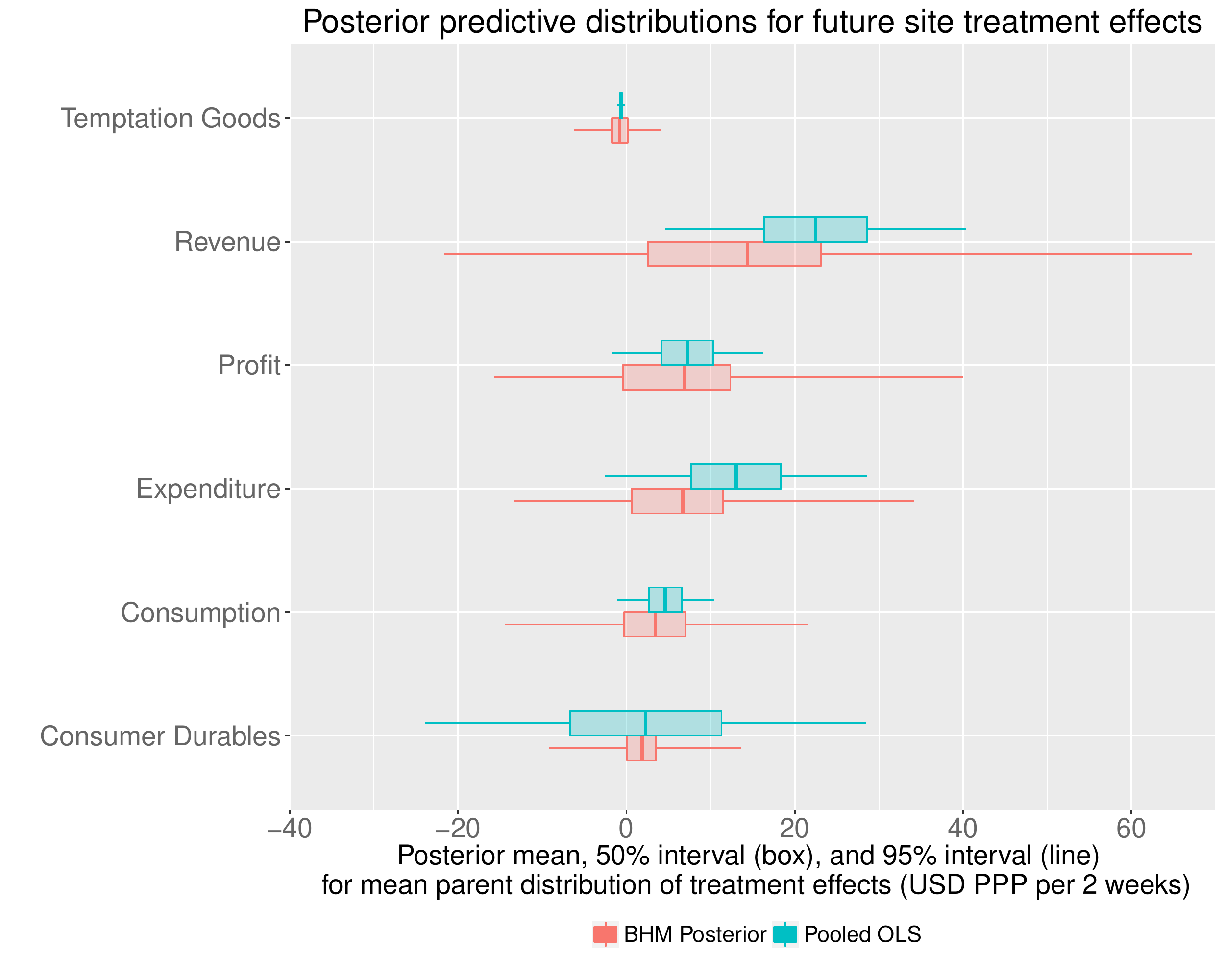}
  \caption{Posterior predictive distributions for the next site, $\tau_{K+1}$} \label{posterior predictive distributions graph}
\end{figure}

\begin{figure}[h!]
  \centering
    \includegraphics[scale=0.6]{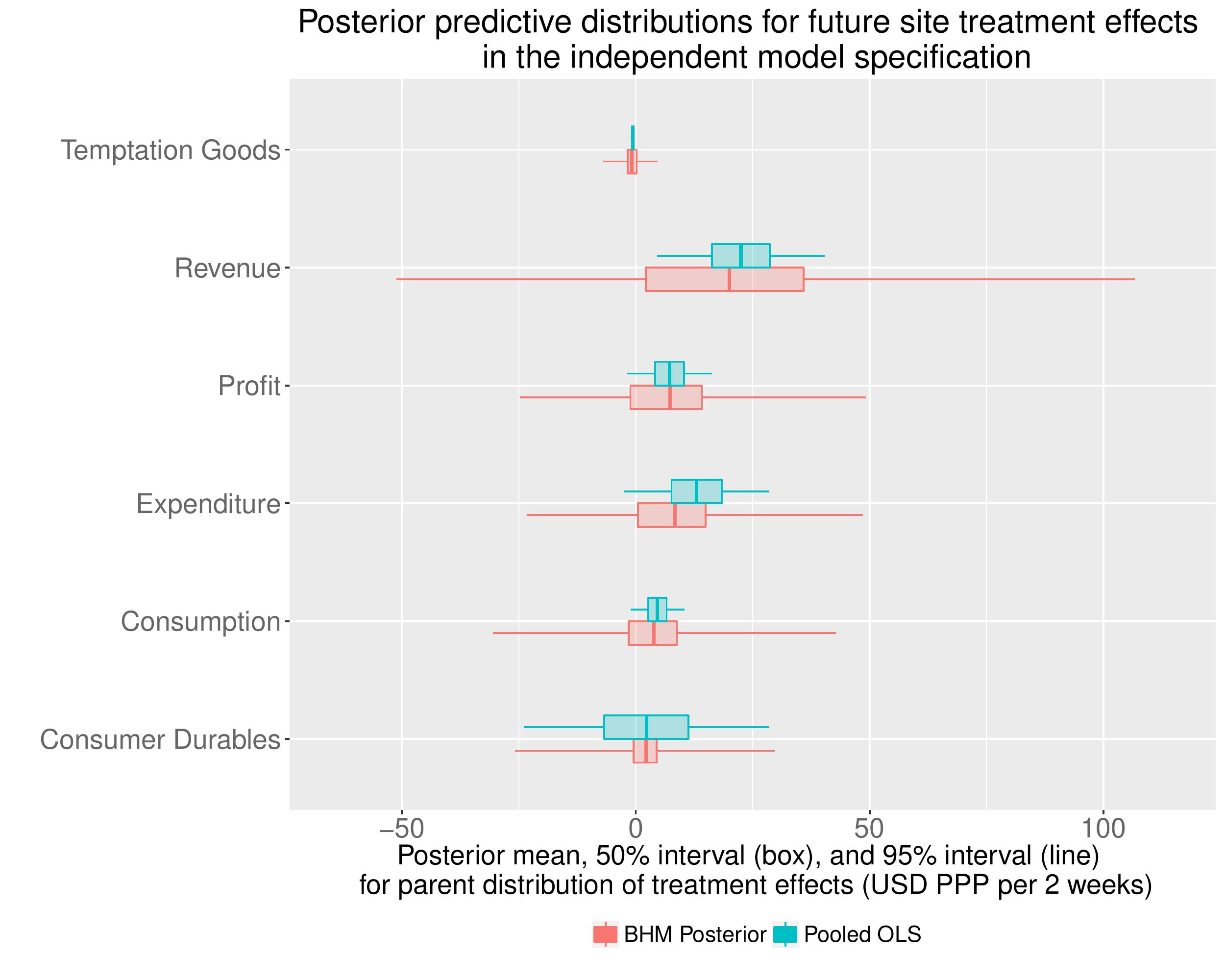}
  \caption{Posterior predictive distributions for the next site, $\tau_{K+1}$ from the independent model} \label{independent posterior predictive distributions graph}
\end{figure}

The posterior predictive distributions of the treatment effects in future exchangeable sites are shown in figure \ref{posterior predictive distributions graph}, with the full-pooling OLS distributions shown for comparison. The results of the independent model are shown in figure \ref{independent posterior predictive distributions graph}, and are somewhat wider than the joint model, because using the observed (typically positive) correlation between the control mean and treatment effect improves fit. The lack of full pooling is evident here in both models, as these intervals are substantially wider than the posterior intervals on $\tau$ - although the pooling that does occur means that they are similar orders of magnitude. These results allow us to make direct probability statements about the future sites effects. For example, the treatment effect for almost all the outcomes has a 25\% chance of realizing in a socially undesirable direction, such as a negative impact on profits and consumer durables spending, or a positive impact on temptation goods spending. More specifically, the next site's treatment effect on profit has a 50\% chance to realize between 0 and 11 USD PPP, a 25\% chance of being negative,  and a 25\% chance of being higher than 11 USD PPP. It has a 95\% chance of realizing between -16 and 40 USD PPP. This is much wider than the OLS estimator's 95\% interval, which spans from -2 to 17 USD PPP, and hence severely underestimates the true uncertainty around the future effect. This is not surprising, as the full pooling model will underestimate this uncertainty in every case except for perfect full pooling and while the microcredit studies exhibit substantial pooling, it is certainly not full pooling. Accounting for this uncertainty is crucial in order to correctly represent the likely outcomes of future interventions to governments and policymakers. 

The finding that the treatment effects of microcredit are reasonably informative for one another differs from the conclusions of Pritchett and Sandefur (2015). This is to be expected, as they analyzed only the results of the no pooling model, which are much more dispersed than the partial pooling model results as shown in figure \ref{tau k posteriors with ols graph}. The Pritchett and Sandefur (2015) analysis is predicated on the idea that there is no common component to the site-specific effects: if there is such a component, the no pooling model produces overdispersed estimates. Because they restrict their analysis to the no pooling model, it is not surprising that they find more dispersion in the treatment effects. My results differ in a similar fashion from the overall assessment of Vivalt (2016) because despite using partial pooling methods, she uses metrics of heterogeneity that include the sampling variation as well. The results of the Bayesian hierarchical model here suggest that much of the apparent dispersion from the no-pooling models is due to sampling variation, and the real underlying heterogeneity is much smaller.

\section{\large understanding Heterogeneity in Treatment Effects}

\subsection{ Household-level Covariates}
Despite the substantial pooling, there remains some heterogeneity in treatment effects across sites. This may be due to a variety of contextual variables that affect households, or variables that affect the sites and studies themselves. Ideally we would like to understand the extent to which conditioning on these covariates can explain the heterogeneity between the observed treatment effects. At the household level, researchers have identified several potentially important covariates such as a household's previous business experience, urban versus rural location, and group versus individual loans (Banerjee et al 2015b, Crepon et al 2015). It is possible that heterogeneous effects across sites are due to differences in composition of household types; alternatively, it is possible that the heterogeneity across sites is generated within one subgroup only. Fitting a fully interacted model with a separate control mean and treatment effect for each of the 8 subgroups implied by all combinations of these 3 variables would be ideal, but this is challenging because two of these variables (loan type and location) did not vary within site at all for 5 of the 7 studies. By contrast, prior business ownership varied within site for all studies except Karlan and Zinman 2011, so this is the natural variable to examine in detail.

\begin{figure}[h!]
  \centering
    \includegraphics[scale=0.6]{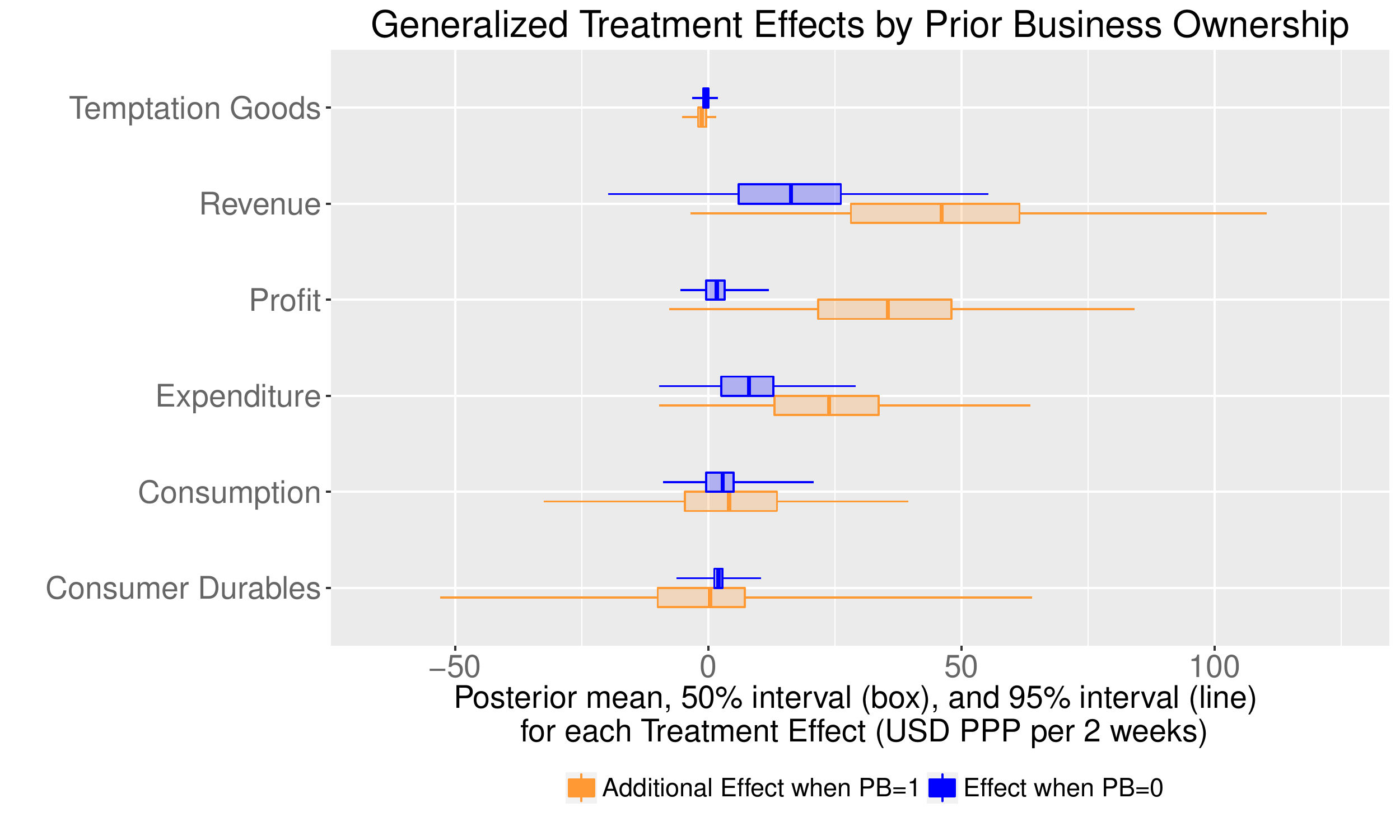}
  \caption{Posterior distributions of $\tau$ for all outcomes split by prior business ownership} \label{posterior split te graph} 
\end{figure}

\begin{figure}[h!]
  \centering
    \includegraphics[scale=0.7]{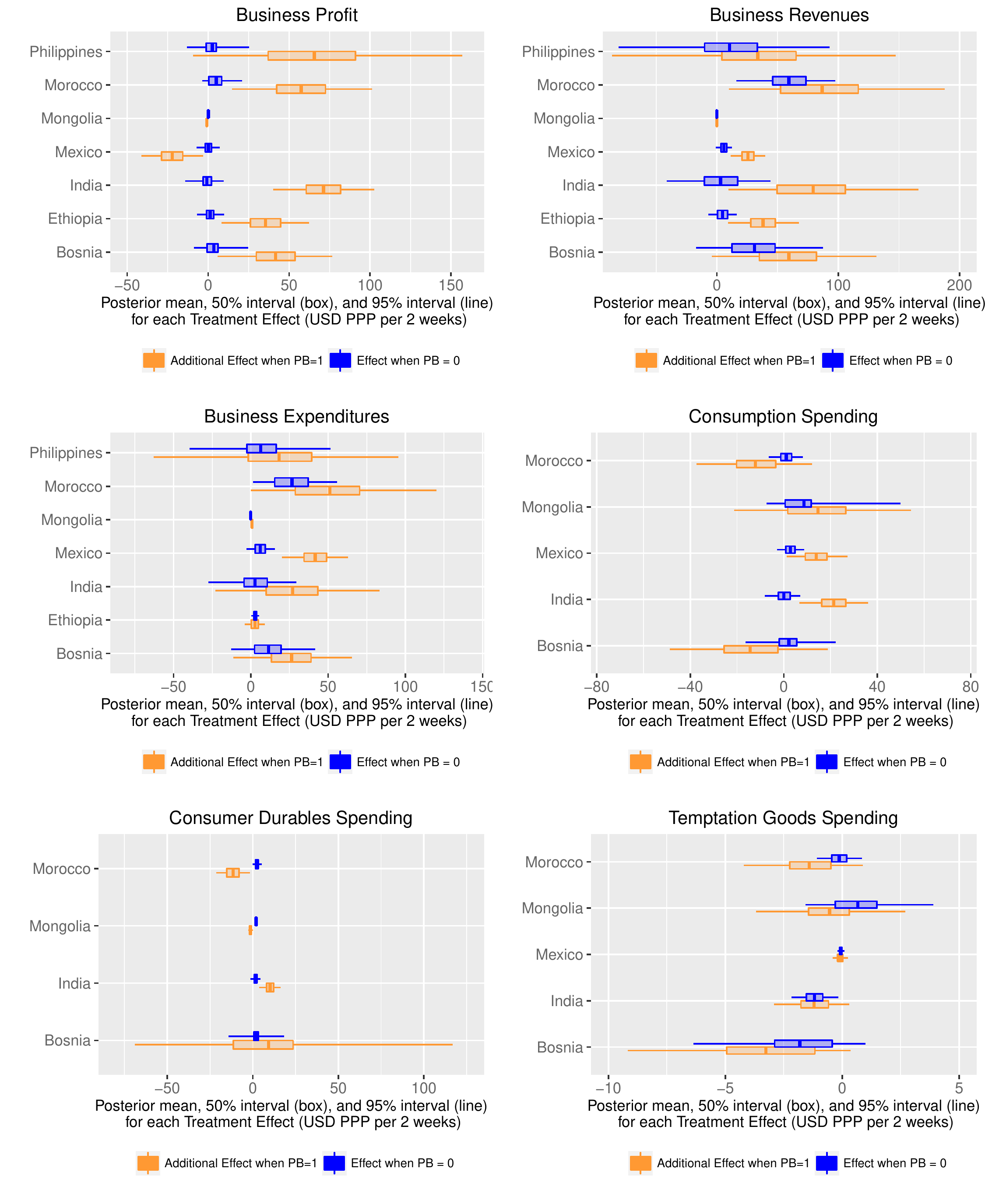}
  \caption{Posterior distributions of $\tau_k$ for all sites and outcomes split by prior business ownership} \label{posterior split te 7 studies graph} 
\end{figure}

I fit an interactions model from equations \ref{interactions model}, enforcing independence in the parent distribution for tractability, and focusing on a single household covariate: a binary indicator on whether the household already operated a business before any microcredit expansion. Denote this variable $PB$, where $PB = 1$ if the household operated a business prior to the microcredit intervention. The results from fitting the fully interacted model with this covariate show that the households where $PB=1$ exhibit much more heterogeneity in treatment effects across sites. Figure \ref{posterior split te graph}  shows the posterior distributions of the general impacts for the two groups, and figure \ref{posterior split te 7 studies graph}  shows the posterior distributions of the impacts for each group in each site. While revenues and expenditures seem to rise for both groups - albeit less for the group with new businesses - only the households with prior business experience appear likely to be making profits.  In fact, the treatment effect on profit is almost exactly zero in every site (panel 1 of figure \ref{posterior split te 7 studies graph}). This suggests that these new business owners are either less productive types or must require a lot of learning or experimentation with their business before they can make profit. 

\begin{figure}[h!]
  \centering
    \includegraphics[scale=0.6]{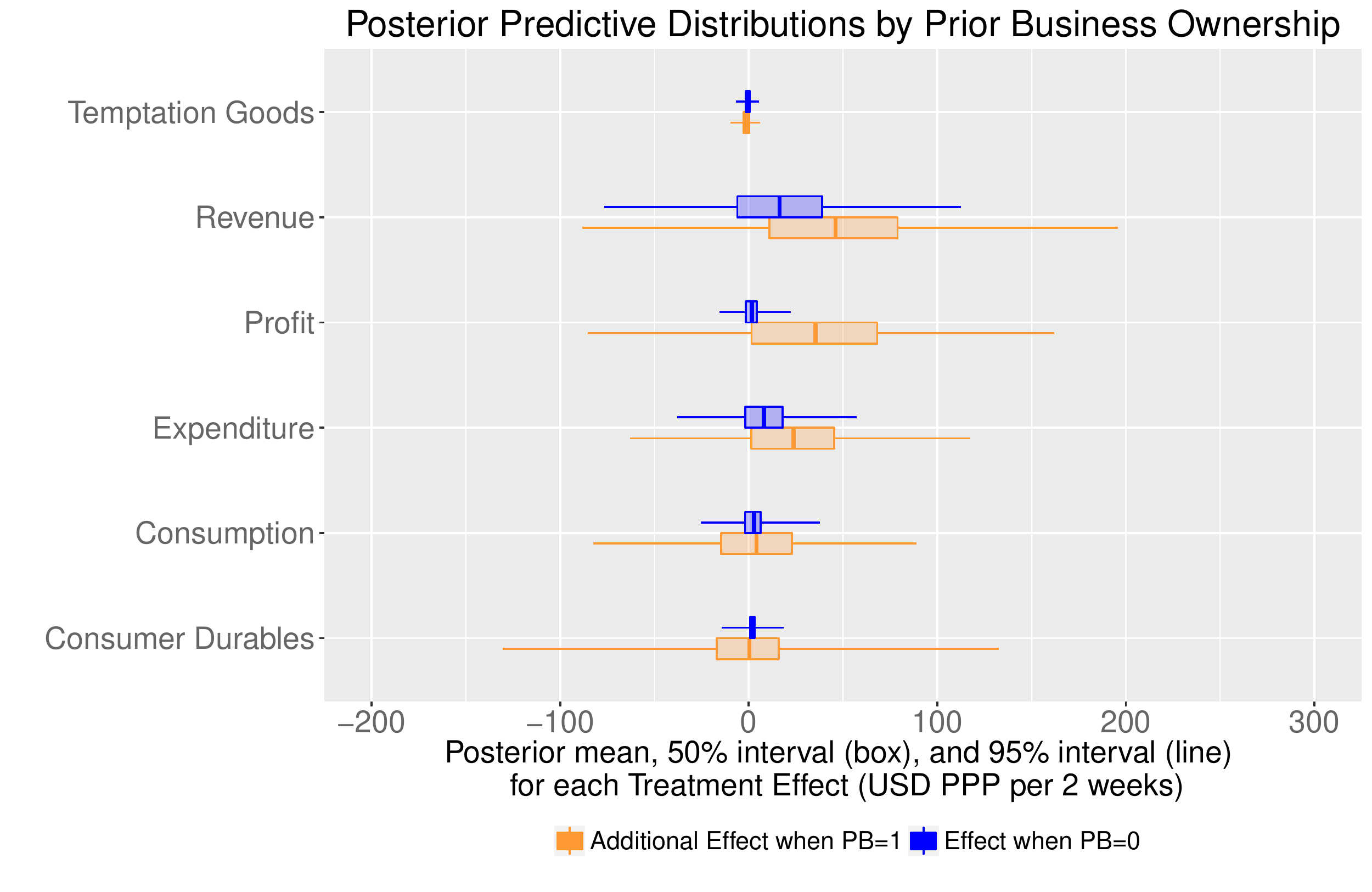}
  \caption{Posterior predictive distributions of $\tau_{K+1}$ split by prior business ownership} \label{posterior predictive split te graph} 
\end{figure}

Overall, there is little cross-site heterogeneity for the households who did not operate a business before the treatment - the only variable which exhibits any variation across site for this group is revenues which is the most volatile outcome in any case. By contrast there is substantial heterogeneity across sites for the households who did operate a previous business - they make up only 27\% of the sample, but they generate most of the cross-site heterogeneity. The posterior means of their additional treatment effects on household profit range from -22 USD PPP to 68 USD PPP per two weeks, although the posteriors themselves are extremely diffuse with very large intervals. The increased heterogeneity in the group with prior business experience is also evident by the much wider posterior predictive distributions for the effects in this group, as shown in figure \ref{posterior predictive split te graph}.

These results illustrate how multi-study analysis can help to combat the problems of searching over subgroups for statistically significant effects. When the researchers who ran the RCT in India (Banerjee et al 2015) checked for this same subgroup, which comprised 29\% of their sample, they found a very large, statistically significant effect here. But the other sites show that this is not always the case - in some cases this subgroup displays a negligible effect, and in others the effect is large and negative. While the general treatment effect for the subgroup of households who had a previous business is indeed much higher than for those without, the posterior distribution of this additional effect is extremely diffuse, and zero is contained in the central 95\% intervals. While there is much more activity in this subgroup, it is not the case that the impact of microcredit is conclusively positive overall even here, as the heterogeneity across sites is much more pronounced in this group.

\subsection{Site-level Covariates}
Important contextual variables also occur at the site or study level, and conditioning on these covariates could explain the remaining heterogeneity between the observed treatment effects. Unfortunately, in the microcredit literature, we face the severe challenge of assessing the role of these covariates with only 7 studies to use as data points. Yet researchers and commentators are already computing these correlations and theorising about the role of contextual variables such as credit market saturation (see for example Wydick 2015). It is highly misleading to focus the analysis on any one contextual variable, because there are many such variables which may be more strongly correlated with the treatment effects if only they were included in the model. To rigorously study the correlations at the site level, the best we can do is regress the treatment effects on all relevant contextual variables using a regularization procedure such as Ridge or Lasso in order to avoid overfitting and detect only the most powerful correlations. 

This exercise can be performed within the Bayesian hierarchical models from section \ref{section: methodology} by modifying the second level of the likelihood to have a regression structure on the mean. Consider a set of $S$ contextual variables, stored in a vector denoted $X_k$ for site $k$. Then we can specify that $\tau_k \sim N(\tau + X_k \beta, \sigma^2_{\tau})$ for all sites, and re-estimate the model. If, as is the case here, the dimensionality of $X_k$ is equal to or larger than $K$, we must enforce sparsity using very strong priors that each element of the slope vector $\beta$ is zero. I standardise all variables to have zero mean and a standard deviation of 1, and use a spherical Ridge prior such that for each row $s$ of the slope vector $\beta$, the prior is $\beta[s] \sim N(0,0.5)$. This implements Bayesian Ridge, in which only the variables with the strongest predictive power for the pattern in $\{\tau_k\}_{k=1}^{K}$ end up with large coefficients, in a similar fashion to the frequentist Ridge procedure (Griffin and Brown, 2013).  For this data I have tested Ridge penalties of size 0.25, 0.5, 1, and 3 with no resulting change in the ordering of coefficients. 

I consider a model with many site-level contextual variables, although this is not exhaustive. In the order in which they appear in the $X_k$ vector, they are: the site's average value of the outcome in the control group, a binary indicator on whether the unit of study randomization was individuals or communities, a binary indicator on whether the MFI targeted female borrowers, the interest rate (APR) at which the MFI in the study usually lends, a microcredit market saturation metric taking integer values from 0-3, a binary indicator on whether the MFI promoted the loans to the public in the treatment areas, a binary indicator on whether the loans were supposed to be collateralized, and the loan size as a percentage of the country's average income per capita. Table \ref{table: contextual variables} displays the values taken by each of these variables in each site, although of course they must be standardized for any sparsity estimation procedure.

\begin{figure}[h!]
  \centering
    \includegraphics[scale=0.8]{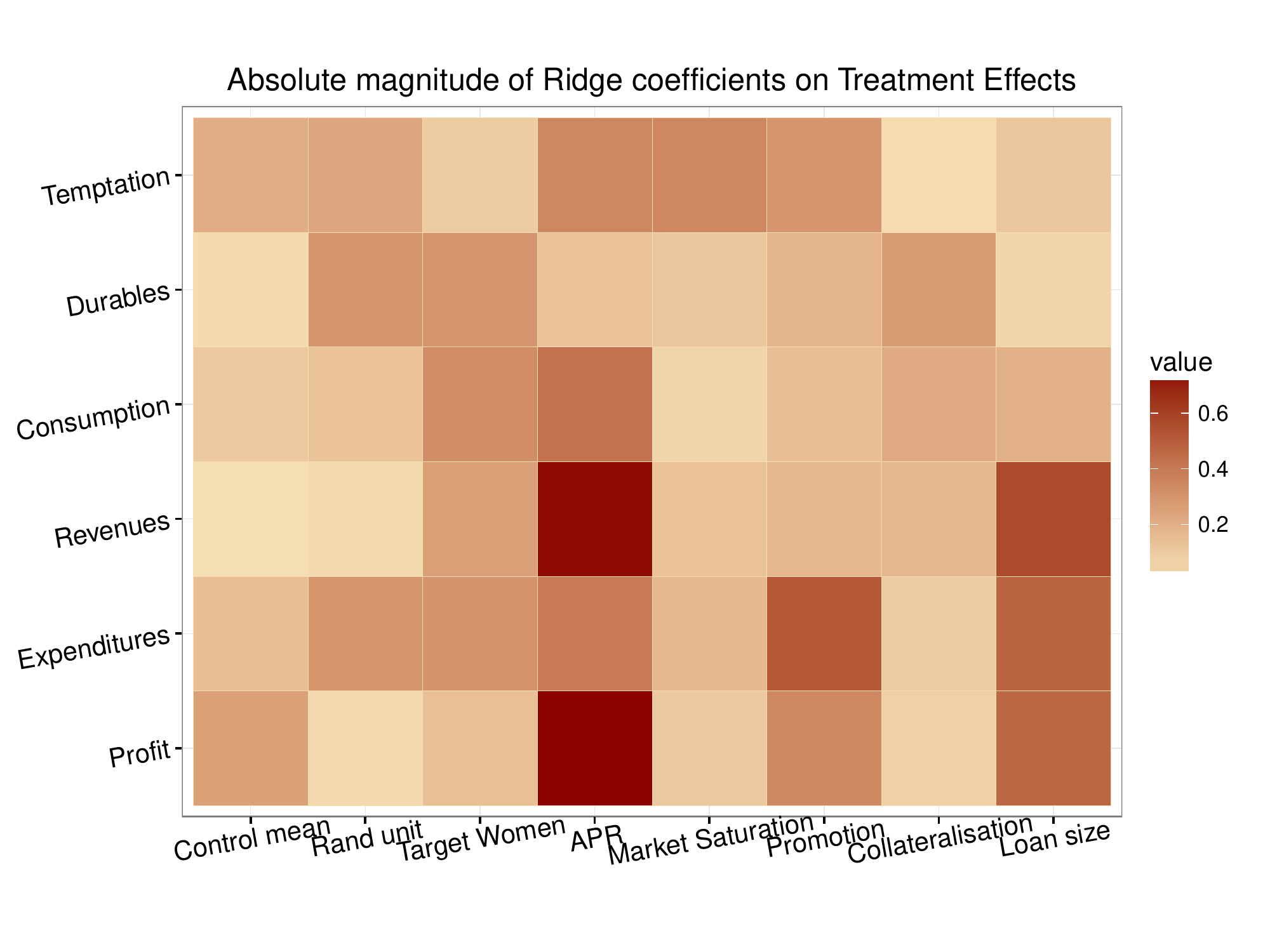}
  \caption{Absolute Magnitude of the Ridge Regression Coefficients for all outcomes and covariates} \label{ridge regression}
\end{figure}

\begin{figure}[h!]
  \centering
    \includegraphics[scale=0.8]{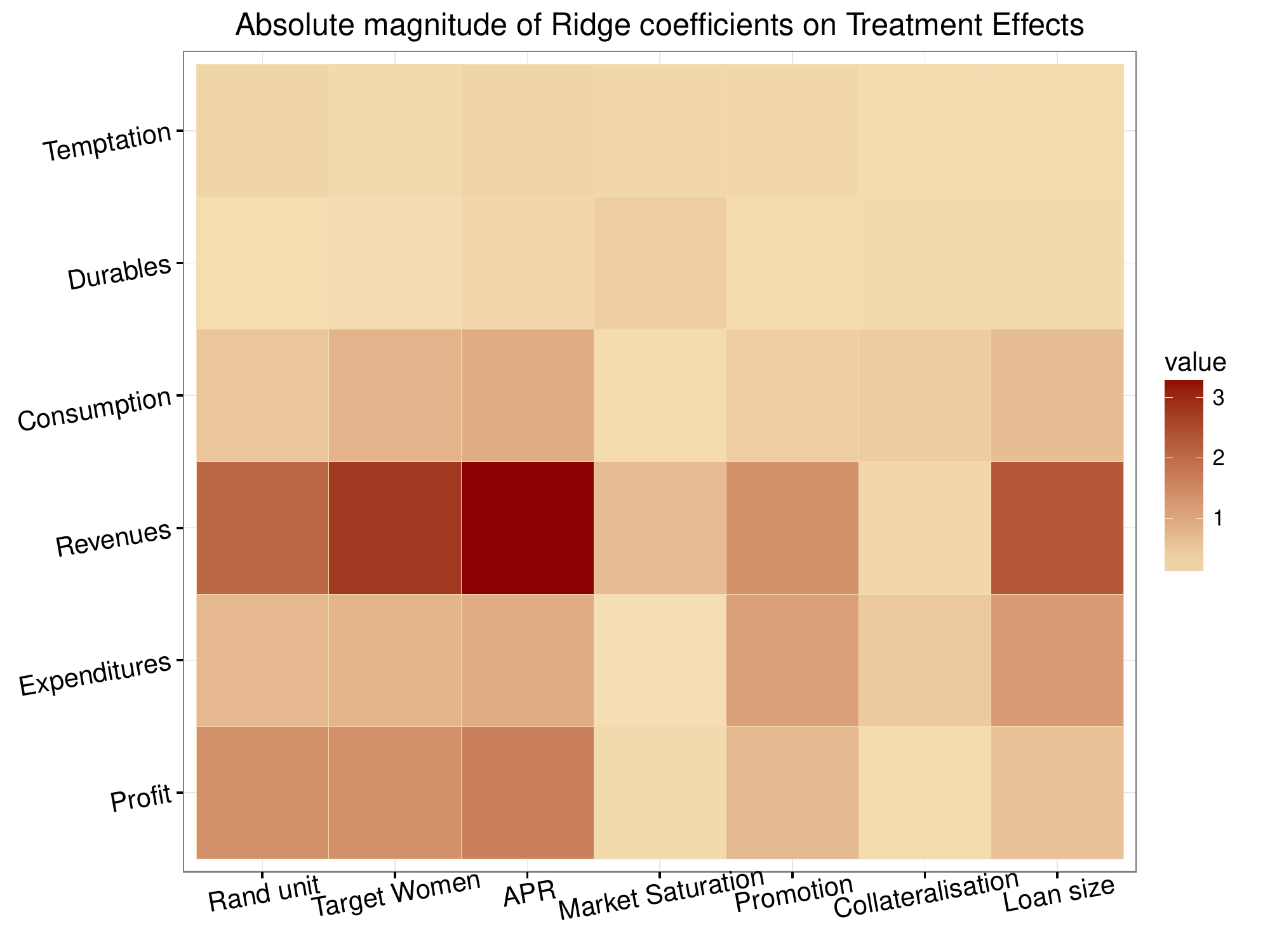}
  \caption{Absolute Magnitude of the Full Bayesian Ridge Regression Coefficients for all outcomes and covariates, omitting the control group mean} \label{ridge regression full bayes}
\end{figure}

The results of a Ridge regression at the study level are shown in figure \ref{ridge regression}, which displays the absolute magnitude of the coefficients on the various contextual variables for each of the 6 outcomes. The larger the magnitude, the more important is the variable as a predictor of the treatment effects for that outcome (Hastie et al, 2009). The figure shows that the most predictive variable is the average interest rate the MFI offers on the loans, followed by the average loan size as a percentage of national household income. The unit of randomization is not highly predictive despite the theoretical case for selection bias in the studies that randomized at the individual level. This could be because the unit of randomization is correlated with the control group mean, although in that case it is still the correct procedure to include both the control mean and the unit of randomization in the Ridge regression. However, to make the point that the unit of randomization is not a strong predictor, I perform a full Bayesian Ridge on both the control mean and the treatment effect, omitting the control mean as an explanatory factor for the treatment effects. I find that the correlation between the interest rate and treatment effects is still much larger than the correlation with the unit of randomization (see figure \ref{ridge regression full bayes}). Thus, these results suggest that the observed heterogeneity is not a function of study design protocols, but rather of genuine economic differences across the settings.

\section{\large Conclusion}


Applying the framework of Bayesian hierarchical models to aggregate the evidence from 7 randomized experiments of expanding access to microcredit has substantially improved our understanding of both the general impact of the intervention and the heterogeneity across contexts. My results suggest that the effect of microcredit access is likely to be positive but small in magnitude relative to control group average levels, and the possibility of a negative impact cannot be ruled out. By contrast, full-pooling methods misleadingly produce ``statistically significant'' results in 2 of the 6 outcomes I study. Standard pooling metrics for the studies indicate on average 60\% pooling on the treatment effects, suggesting that the site-specific effects are reasonably informative and externally valid for each other and for the general case. Further analysis incorporating household covariates shows that the cross-study heterogeneity is almost entirely generated by heterogeneous effects for the 27\% households who previously operated businesses before microcredit expansion. Assessing the role of site-specific covariates using a Ridge procedure indicates that economic variables, in particular the interest rate on the microloans and the loan size, have the strongest correlation with the treatment effects.  

Although these results provide us with a deep understanding of the evidence on the impact of expanding access to microcredit, some unresolved questions remain. One major concern is that microcredit could affect household or village welfare without affecting the average outcomes: perhaps households use these loans to manage risk, or reallocate investment or consumption towards more durable or lumpy goods. There could also be heterogeneous effects of microcredit within villages - this is why the original papers reported quantile treatment effects at each decile (see for example Banerjee et al 2015). However, aggregating quantile treatment effects using Bayesian hierarchical models would require partial pooling while respecting monotonicity of the conditional quantiles. Such methods have not yet been developed,  but this should be the focus of future research. Another issue I have not addressed in this paper is that of combining experimental and observational data. I chose instead to focus on the experimental studies, in which the estimated effects are more credibly causal. However, building a structure that can account for the differing strength of the exclusion restrictions across studies into the Bayesian hierarchical framework should also be a priority for future work.

Nevertheless, the results presented here do illustrate the importance of rigorously aggregating evidence across heterogeneous settings in development economics. My results differ substantially from the conclusions drawn in informal review articles such as Banerjee et al (2015a) and previous attempts to formally aggregate evidence on microcredit such as Pritchett and Sandefur (2015) and Vivalt (2016) which failed to separate sampling variation in the estimates from genuine underlying heterogeneity. The results of the Bayesian hierarchical model also differ from the results of more common full-pooling aggregation methods, which in 2 of the 6 outcomes would misleadingly detect a ``statistically significant'' result on average. While the econometric issues involved in evidence aggregation may seem abstract or technical, they can have a major impact on the conclusions we draw from the evidence we have. If development economists seek to produce reliable bodies of generalizable evidence for policymakers to use, then aggregating results and assessing external validity via Bayesian hierarchical models should become a core part of the research-to-policy pipeline.

\newpage

\clearpage

\clearpage

\begin{table}[ht]
\centering
\small
\title{Contextual Variables (Pre-Standardization)}
\begin{tabular}{rrrrrrrr}
  \hline
 & Rand unit & Women & APR & Saturation & Promotion & Collateral & Loan size \\ 
  \hline
Mexico (Angelucci) & 0 & 1 & 100.00 & 2& 1 & 0 & 6.00 \\ 
  Mongolia (Attanasio) & 0 & 1 & 120.00 & 1 & 0 & 1 & 36.00 \\ 
  Bosnia (Augsberg) & 1 & 0 & 22.00 & 2 & 0& 1 & 9.00 \\ 
  India (Banerjee) & 0 & 1 & 24.00 & 3 & 0& 0 & 22.00 \\ 
  Morocco (Crepon) & 0 & 0 & 13.50 & 0 & 1& 0& 21.00 \\ 
  Philippines (Karlan) & 1 & 0 & 63.00 & 1& 0 & 0 & 24.10 \\ 
  Ethiopia (Tarozzi) & 0 & 0 & 12.00 & 1 & 0 & 0 & 118.00 \\ 

   \hline
\end{tabular}
\caption{Contextual Variables: Unit of randomization (1 = individual, 0 = community), Women (1= MFI targets women, 0 = otherwise), APR (annual interest rate), Saturation metric (3 = highly saturated, 0 = no other microlenders operate), Promotion (1 = MFI advertised itself in area, 0 = no advertising), Collateral (1 = MFI required collateral, 0 = no collateral required), Loan size (percentage of mean national income)} \label{table: contextual variables}
\end{table}

\clearpage
\appendix
\appendixpage
\section{ Tables of Marginal Posteriors for the Main Specification}

\begin{table}[ht]
\centering
\title{Marginal Posteriors of the Joint Model: Profit} 
\begin{tabular}{rrrrrrrrrrr}
  \hline
	&	mean	&	2.5\%	&	25\%	&	50\%	&	75\%	&	97.5\%	\\
	\hline
$\mu$	&	94.81	&	-20.6	&	59.18	&	93.57	&	129.76	&	213.79	\\
$\tau$	&	6.81	&	-3.03	&	1.82	&	5.37	&	10.38	&	24.49	\\
$\mu_1$	&	12.52	&	5.3	&	9.91	&	12.47	&	15.11	&	20.06	\\
$\tau_1$	&	-0.77	&	-10.96	&	-3.68	&	-0.48	&	2.36	&	8.18	\\
$\mu_2$	&	-0.68	&	-1.07	&	-0.81	&	-0.68	&	-0.55	&	-0.3	\\
$\tau_2$	&	-0.34	&	-0.77	&	-0.49	&	-0.33	&	-0.18	&	0.1	\\
$\mu_3$	&	111.24	&	87.97	&	103.71	&	111.56	&	119.07	&	132.83	\\
$\tau_3$	&	11.27	&	-4.6	&	2.66	&	8.71	&	17.52	&	39.93	\\
$\mu_4$	&	34.04	&	19.15	&	29.19	&	34.34	&	39.1	&	47.52	\\
$\tau_4$	&	8.09	&	-4.54	&	2.09	&	6.77	&	12.98	&	26.9	\\
$\mu_5$	&	85.38	&	71.05	&	80.57	&	85.45	&	90.26	&	99.56	\\
$\tau_5$	&	9.04	&	-3.82	&	2.63	&	7.7	&	14.32	&	28.32	\\
$\mu_6$	&	405.16	&	329.58	&	381.67	&	406.12	&	430.37	&	472.85	\\
$\tau_6$	&	15.01	&	-11.79	&	1.79	&	9.44	&	22.88	&	71.11	\\
$\mu_7$	&	14.1	&	4.13	&	10.81	&	14.21	&	17.44	&	23.46	\\
$\tau_7$	&	5.29	&	-4.98	&	1.22	&	4.6	&	9.03	&	18.24	\\
$\sigma_{y1}$	&	378.27	&	374.27	&	376.82	&	378.26	&	379.69	&	382.39	\\
$\sigma_{y2}$	&	3.08	&	2.94	&	3.03	&	3.07	&	3.12	&	3.22	\\
$\sigma_{y3}$	&	342.13	&	328.92	&	337.22	&	342.07	&	346.83	&	355.98	\\
$\sigma_{y4}$	&	489.49	&	481.24	&	486.58	&	489.46	&	492.36	&	497.86	\\
$\sigma_{y5}$	&	422.74	&	414.91	&	420.04	&	422.72	&	425.45	&	430.71	\\
$\sigma_{y6}$	&	1041.14	&	999.38	&	1026.14	&	1040.63	&	1056.03	&	1084.79	\\
$\sigma_{y7}$	&	220.16	&	214.7	&	218.26	&	220.1	&	222.03	&	225.82	\\
$\Omega_{11}$	&	1	&	1	&	1	&	1	&	1	&	1	\\
$\Omega_{12}$	&	0.2	&	-0.58	&	-0.07	&	0.23	&	0.5	&	0.84	\\
$\Omega_{21}$	&	0.2	&	-0.58	&	-0.07	&	0.23	&	0.5	&	0.84	\\
$\Omega_{22}$	&	1	&	1	&	1	&	1	&	1	&	1	\\
$\theta_1$	&	146.64	&	80.43	&	112.58	&	137.02	&	169.54	&	268.84	\\
$\theta_2$	&	9.35	&	1.66	&	4.39	&	7.67	&	12.26	&	27.38	\\
$V_{11}$	&	23982.21	&	6469.65	&	12674.01	&	18774.8	&	28745.47	&	72274.9	\\
$V_{12}$	&	319.52	&	-838.72	&	-51.43	&	181.18	&	570.83	&	2151.07	\\
$V_{21}$	&	319.52	&	-838.72	&	-51.43	&	181.18	&	570.83	&	2151.07	\\
$V_{22}$	&	136.25	&	2.77	&	19.29	&	58.89	&	150.3	&	749.71	\\
$\mu_{K+1}$	&	93.75	&	-234.71	&	-4.12	&	92.62	&	191.91	&	424.15	\\
$\tau_{K+1}$	&	6.89	&	-15.7	&	-0.41	&	4.49	&	12.35	&	40.04	 \\

   \hline
\end{tabular}
\end{table}

\begin{table}[ht]
\centering
\title{Marginal Posteriors of the Joint Model: Revenues}
\begin{tabular}{rrrrrrrrrrr}
  \hline
	&	mean	&	2.5\%	&	25\%	&	50\%	&	75\%	&	97.5\%	\\
\hline													
$\mu$	&	306.17	&	-87.46	&	185.89	&	307.29	&	428.43	&	696.44	\\
$\tau$	&	14.45	&	-1.4	&	6.58	&	12.13	&	19.93	&	43.53	\\
$\mu_1$	&	45.03	&	39.02	&	42.92	&	44.99	&	47.09	&	51.43	\\
$\tau_1$	&	9.36	&	0.73	&	6.57	&	9.41	&	12.25	&	17.59	\\
$\mu_2$	&	1.05	&	0.76	&	0.95	&	1.05	&	1.16	&	1.35	\\
$\tau_2$	&	-0.07	&	-0.41	&	-0.18	&	-0.07	&	0.05	&	0.28	\\
$\mu_3$	&	185.16	&	142.49	&	170.94	&	185.42	&	199.26	&	227.07	\\
$\tau_3$	&	21.33	&	-4.43	&	7.58	&	16.67	&	30.76	&	71.28	\\
$\mu_4$	&	208.88	&	172.23	&	196.54	&	209.04	&	221.51	&	244.39	\\
$\tau_4$	&	14.78	&	-10.36	&	4.54	&	12.19	&	22.73	&	51.66	\\
$\mu_5$	&	332.34	&	300.85	&	322.4	&	332.87	&	342.8	&	360.67	\\
$\tau_5$	&	25.04	&	-1.25	&	10.71	&	21.05	&	36.02	&	70.68	\\
$\mu_6$	&	1432.05	&	1281.01	&	1382.19	&	1432.74	&	1483.16	&	1578.58	\\
$\tau_6$	&	20.64	&	-24.63	&	3.55	&	14.66	&	31.87	&	98.07	\\
$\mu_7$	&	26.19	&	15.49	&	22.54	&	26.3	&	29.89	&	36.69	\\
$\tau_7$	&	10.26	&	-2.59	&	5.4	&	9.99	&	14.87	&	24.6	\\
$\sigma_{y1}$	&	276.33	&	273.38	&	275.29	&	276.34	&	277.37	&	279.33	\\
$\sigma_{y2}$	&	2.42	&	2.31	&	2.38	&	2.42	&	2.45	&	2.53	\\
$\sigma_{y3}$	&	637.45	&	612.49	&	628.43	&	637.14	&	646.2	&	663.8	\\
$\sigma_{y4}$	&	1372.56	&	1349.97	&	1364.52	&	1372.38	&	1380.68	&	1395.55	\\
$\sigma_{y5}$	&	891.33	&	874.85	&	885.55	&	891.32	&	897.01	&	908.61	\\
$\sigma_{y6}$	&	2374.1	&	2279.02	&	2338.83	&	2373.71	&	2407.68	&	2475.84	\\
$\sigma_{y7}$	&	225.62	&	220.06	&	223.73	&	225.58	&	227.5	&	231.14	\\
$\Omega_{11}$	&	1	&	1	&	1	&	1	&	1	&	1	\\
$\Omega_{12}$	&	0.13	&	-0.59	&	-0.14	&	0.15	&	0.41	&	0.78	\\
$\Omega_{21}$	&	0.13	&	-0.59	&	-0.14	&	0.15	&	0.41	&	0.78	\\
$\Omega_{22}$	&	1	&	1	&	1	&	1	&	1	&	1	\\
$\theta_1$	&	521.33	&	300.62	&	407.33	&	488.99	&	595.67	&	942.13	\\
$\theta_2$	&	14.98	&	3.02	&	7.4	&	12.03	&	19.29	&	43.64	\\
$V_{11}$	&	300053.45	&	90370.01	&	165914.61	&	239111	&	354826.33	&	887616.37	\\
$V_{12}$	&	1050.6	&	-6150.13	&	-677.52	&	651.58	&	2385.41	&	10298.03	\\
$V_{21}$	&	1050.6	&	-6150.13	&	-677.52	&	651.58	&	2385.41	&	10298.03	\\
$V_{22}$	&	343.41	&	9.1	&	54.73	&	144.83	&	372.08	&	1904.23	\\
$\mu_{K+1}$	&	304.8	&	-873.74	&	-45.89	&	302.79	&	657.91	&	1461.35	\\
$\tau_{K+1}$	&	14.4	&	-21.61	&	2.58	&	11.04	&	23.07	&	67.18	\\

   \hline
\end{tabular}
\end{table}

\begin{table}[ht]
\centering
\title{Marginal Posteriors of the Joint Model: Expenditures} 
\begin{tabular}{rrrrrrrrrrr}
  \hline
	&	mean	&	2.5\%	&	25\%	&	50\%	&	75\%	&	97.5\%	\\
\hline													
$\mu$	&	212.58	&	-83.92	&	124.88	&	213.75	&	298.43	&	503.32	\\
$\tau$	&	6.72	&	-2.3	&	2.57	&	5.54	&	9.7	&	22.07	\\
$\mu_1$	&	34.84	&	26.51	&	32.13	&	34.93	&	37.63	&	42.63	\\
$\tau_1$	&	8.83	&	-0.13	&	4.57	&	8.36	&	12.58	&	20.71	\\
$\mu_2$	&	1.73	&	1.19	&	1.54	&	1.73	&	1.91	&	2.26	\\
$\tau_2$	&	0.29	&	-0.34	&	0.09	&	0.29	&	0.5	&	0.9	\\
$\mu_3$	&	69.45	&	46.4	&	61.55	&	69.4	&	77.39	&	91.62	\\
$\tau_3$	&	9.2	&	-4.56	&	2.64	&	7.03	&	13.93	&	33.53	\\
$\mu_4$	&	172.12	&	141.77	&	161.64	&	172.16	&	182.83	&	201.92	\\
$\tau_4$	&	6.48	&	-9.89	&	1.06	&	5.11	&	10.89	&	28.58	\\
$\mu_5$	&	194.17	&	174.09	&	187.73	&	194.6	&	200.84	&	212.92	\\
$\tau_5$	&	10.39	&	-3.38	&	3.37	&	8.1	&	15.47	&	35.13	\\
$\mu_6$	&	1033.17	&	910.58	&	991.14	&	1032.69	&	1075.59	&	1156.42	\\
$\tau_6$	&	8.59	&	-19.09	&	-0.03	&	6.12	&	15.23	&	47.49	\\
$\mu_7$	&	12.8	&	10.4	&	12.01	&	12.8	&	13.62	&	15.17	\\
$\tau_7$	&	3.57	&	0.38	&	2.4	&	3.54	&	4.7	&	6.94	\\
$\sigma_{y1}$	&	392.84	&	388.59	&	391.39	&	392.88	&	394.34	&	396.98	\\
$\sigma_{y2}$	&	4.37	&	4.18	&	4.3	&	4.37	&	4.44	&	4.57	\\
$\sigma_{y3}$	&	365.97	&	352.03	&	360.83	&	365.88	&	370.99	&	381.11	\\
$\sigma_{y4}$	&	1208.67	&	1188.84	&	1201.45	&	1208.47	&	1215.66	&	1229.18	\\
$\sigma_{y5}$	&	639.14	&	627.49	&	635.03	&	639.21	&	643.13	&	651.03	\\
$\sigma_{y6}$	&	1984.82	&	1904.8	&	1955.37	&	1983.99	&	2013.98	&	2069.98	\\
$\sigma_{y7}$	&	48.42	&	47.22	&	48	&	48.41	&	48.84	&	49.68	\\
$\Omega_{11}$	&	1	&	1	&	1	&	1	&	1	&	1	\\
$\Omega_{12}$	&	0.07	&	-0.67	&	-0.21	&	0.07	&	0.35	&	0.76	\\
$\Omega_{21}$	&	0.07	&	-0.67	&	-0.21	&	0.07	&	0.35	&	0.76	\\
$\Omega_{22}$	&	1	&	1	&	1	&	1	&	1	&	1	\\
$\theta_1$	&	380.37	&	221.21	&	296.55	&	353.68	&	432.96	&	697.09	\\
$\theta_2$	&	8.13	&	1.44	&	3.98	&	6.67	&	10.58	&	23.5	\\
$V_{11}$	&	160925.99	&	48934.79	&	87943.11	&	125090.07	&	187458.5	&	485934.76	\\
$V_{12}$	&	228.66	&	-2939.41	&	-416.46	&	111.14	&	800.89	&	3893.86	\\
$V_{21}$	&	228.66	&	-2939.41	&	-416.46	&	111.14	&	800.89	&	3893.86	\\
$V_{22}$	&	101.66	&	2.08	&	15.85	&	44.47	&	111.89	&	552.33	\\
$\mu_{K+1}$	&	214.85	&	-631.34	&	-41.05	&	211.81	&	468.59	&	1071.89	\\
$\tau_{K+1}$	&	6.69	&	-13.36	&	0.64	&	5.06	&	11.46	&	34.12	\\

   \hline
\end{tabular}
\end{table}

\begin{table}[ht]
\centering
\title{Marginal Posteriors of the Joint Model: Consumption} 
\begin{tabular}{rrrrrrrrrrr}
  \hline
	&	mean	&	2.5\%	&	25\%	&	50\%	&	75\%	&	97.5\%	\\
\hline													
$\mu$	&	281.8	&	226.11	&	266.44	&	281.99	&	296.82	&	340.51	\\
$\tau$	&	3.44	&	-6.28	&	0.82	&	3.46	&	5.93	&	13.21	\\
$\mu_1$	&	299.48	&	294.26	&	297.74	&	299.47	&	301.15	&	304.64	\\
$\tau_1$	&	4.48	&	-2.06	&	2.19	&	4.44	&	6.64	&	11.49	\\
$\mu_2$	&	310.69	&	280.94	&	300.57	&	310.55	&	320.78	&	339.16	\\
$\tau_2$	&	5.57	&	-7.4	&	1.07	&	4.48	&	8.25	&	27	\\
$\mu_3$	&	195.89	&	174.11	&	188.51	&	196.06	&	203.37	&	217.33	\\
$\tau_3$	&	1.73	&	-17.94	&	-1.67	&	2.59	&	6.17	&	16.47	\\
$\mu_4$	&	276.79	&	270.01	&	274.52	&	276.83	&	279.02	&	283.32	\\
$\tau_4$	&	3.82	&	-4.09	&	1.2	&	3.75	&	6.3	&	12.17	\\
$\mu_5$	&	325.15	&	317.29	&	322.67	&	325.06	&	327.69	&	332.77	\\
$\tau_5$	&	1.44	&	-8.71	&	-1.26	&	1.8	&	4.58	&	9.59	\\
$\sigma_{y1}$	&	262.18	&	259.21	&	261.2	&	262.19	&	263.14	&	265.09	\\
$\sigma_{y2}$	&	444.06	&	424.74	&	437.35	&	444.02	&	450.72	&	463.98	\\
$\sigma_{y3}$	&	302.02	&	289.19	&	297.34	&	301.81	&	306.66	&	315.31	\\
$\sigma_{y4}$	&	226.13	&	222.43	&	224.8	&	226.12	&	227.41	&	229.97	\\
$\sigma_{y5}$	&	222.94	&	218.82	&	221.52	&	222.93	&	224.37	&	227.08	\\
$\Omega_{11}$	&	1	&	1	&	1	&	1	&	1	&	1	\\
$\Omega_{12}$	&	0.02	&	-0.69	&	-0.27	&	0.02	&	0.29	&	0.74	\\
$\Omega_{21}$	&	0.02	&	-0.69	&	-0.27	&	0.02	&	0.29	&	0.74	\\
$\Omega_{22}$	&	1	&	1	&	1	&	1	&	1	&	1	\\
$\theta_1$	&	56.02	&	28.42	&	41.42	&	51.23	&	63.98	&	113.85	\\
$\theta_2$	&	5.55	&	0.75	&	2.12	&	4.07	&	7.15	&	19.07	\\
$V_{11}$	&	3677.94	&	807.93	&	1715.51	&	2624.56	&	4093.89	&	12961.37	\\
$V_{12}$	&	7.79	&	-317.28	&	-45.19	&	2.4	&	55.06	&	363.04	\\
$V_{21}$	&	7.79	&	-317.28	&	-45.19	&	2.4	&	55.06	&	363.04	\\
$V_{22}$	&	60.38	&	0.56	&	4.5	&	16.53	&	51.14	&	363.68	\\
$\mu_{K+1}$	&	281.91	&	146.46	&	244.37	&	281.64	&	319.68	&	415.24	\\
$\tau_{K+1}$	&	3.45	&	-14.45	&	-0.28	&	3.49	&	7.03	&	21.56	\\

   \hline
\end{tabular}
\end{table}

\begin{table}[ht]
\centering
\title{Marginal Posteriors of the Joint Model: Consumer Durables} 
\begin{tabular}{rrrrrrrrrrr}
  \hline
	&	mean	&	2.5\%	&	25\%	&	50\%	&	75\%	&	97.5\%	\\
\hline													
$\mu$	&	274.31	&	-317.48	&	107.71	&	287.34	&	459.81	&	818.9	\\
$\tau$	&	1.83	&	-3.9	&	0.67	&	1.6	&	2.88	&	8.29	\\
$\mu_1$	&	5.34	&	4.52	&	5.09	&	5.34	&	5.62	&	6.15	\\
$\tau_1$	&	1.09	&	0.18	&	0.76	&	1.08	&	1.41	&	2.06	\\
$\mu_2$	&	1114.28	&	913.26	&	1049.05	&	1116.37	&	1179.95	&	1303.42	\\
$\tau_2$	&	1.87	&	-12.06	&	-0.16	&	1.49	&	3.89	&	16.54	\\
$\mu_3$	&	24.65	&	21.45	&	23.67	&	24.76	&	25.71	&	27.42	\\
$\tau_3$	&	2.89	&	-0.33	&	1.43	&	2.61	&	4.15	&	7.28	\\
$\mu_4$	&	6.43	&	3.71	&	5.56	&	6.41	&	7.34	&	9.18	\\
$\tau_4$	&	1.52	&	-1.89	&	0.51	&	1.44	&	2.53	&	5.1	\\
$\sigma_{y1}$	&	6.83	&	6.53	&	6.73	&	6.83	&	6.94	&	7.15	\\
$\sigma_{y2}$	&	2973.79	&	2853.39	&	2927.51	&	2972.68	&	3018.15	&	3113.24	\\
$\sigma_{y3}$	&	92.59	&	91.06	&	92.08	&	92.58	&	93.1	&	94.09	\\
$\sigma_{y4}$	&	83.56	&	82.11	&	82.99	&	83.55	&	84.12	&	85.17	\\
$\Omega_{11}$	&	1	&	1	&	1	&	1	&	1	&	1	\\
$\Omega_{12}$	&	0.02	&	-0.72	&	-0.26	&	0.02	&	0.3	&	0.74	\\
$\Omega_{21}$	&	0.02	&	-0.72	&	-0.26	&	0.02	&	0.3	&	0.74	\\
$\Omega_{22}$	&	1	&	1	&	1	&	1	&	1	&	1	\\
$\theta_1$	&	625.84	&	277.02	&	407.24	&	519.94	&	684.39	&	2122.43	\\
$\theta_2$	&	3.36	&	0.31	&	1.01	&	2.21	&	4.25	&	13.23	\\
$V_{11}$	&	562199.05	&	76741	&	165846.9	&	270342.64	&	468388.27	&	4504714.78	\\
$V_{12}$	&	18.6	&	-2507.6	&	-275.11	&	11.62	&	327.21	&	2526.88	\\
$V_{21}$	&	18.6	&	-2507.6	&	-275.11	&	11.62	&	327.21	&	2526.88	\\
$V_{22}$	&	26.89	&	0.1	&	1.02	&	4.88	&	18.02	&	174.98	\\
$\mu_{K+1}$	&	271.94	&	-1287.38	&	-116.5	&	275.42	&	669.17	&	1787.86	\\
$\tau_{K+1}$	&	1.85	&	-9.21	&	0.11	&	1.52	&	3.53	&	13.67	\\

   \hline
\end{tabular}
\end{table}

\begin{table}[ht]
\centering
\title{Marginal Posteriors of the Joint Model: Temptation Goods} 
\begin{tabular}{rrrrrrrrrrr}
  \hline
	&	mean	&	2.5\%	&	25\%	&	50\%	&	75\%	&	97.5\%	\\
\hline													
$\mu$	&	18.64	&	3.9	&	14.51	&	18.7	&	22.81	&	33.09	\\
$\tau$	&	-0.79	&	-3.33	&	-1.26	&	-0.7	&	-0.22	&	1.28	\\
$\mu_1$	&	4.69	&	4.56	&	4.64	&	4.69	&	4.73	&	4.82	\\
$\tau_1$	&	-0.09	&	-0.27	&	-0.15	&	-0.09	&	-0.02	&	0.09	\\
$\mu_2$	&	8.28	&	5.42	&	7.44	&	8.33	&	9.18	&	10.8	\\
$\tau_2$	&	0.03	&	-2.4	&	-0.78	&	-0.11	&	0.72	&	3.22	\\
$\mu_3$	&	31.7	&	28.44	&	30.53	&	31.61	&	32.81	&	35.36	\\
$\tau_3$	&	-1.99	&	-6.8	&	-2.97	&	-1.58	&	-0.64	&	0.71	\\
$\mu_4$	&	15.79	&	14.92	&	15.5	&	15.8	&	16.08	&	16.63	\\
$\tau_4$	&	-1.35	&	-2.54	&	-1.77	&	-1.36	&	-0.94	&	-0.1	\\
$\mu_5$	&	32.03	&	31.08	&	31.71	&	32.04	&	32.35	&	32.96	\\
$\tau_5$	&	-0.53	&	-1.81	&	-0.96	&	-0.52	&	-0.11	&	0.74	\\
$\sigma_{y1}$	&	6.05	&	5.99	&	6.03	&	6.05	&	6.08	&	6.12	\\
$\sigma_{y2}$	&	28.98	&	27.72	&	28.54	&	28.96	&	29.41	&	30.32	\\
$\sigma_{y3}$	&	44.28	&	42.41	&	43.57	&	44.25	&	44.95	&	46.33	\\
$\sigma_{y4}$	&	23.76	&	23.36	&	23.62	&	23.76	&	23.9	&	24.16	\\
$\sigma_{y5}$	&	26.79	&	26.3	&	26.62	&	26.79	&	26.96	&	27.29	\\
$\Omega_{11}$	&	1	&	1	&	1	&	1	&	1	&	1	\\
$\Omega_{12}$	&	-0.16	&	-0.75	&	-0.42	&	-0.18	&	0.08	&	0.53	\\
$\Omega_{21}$	&	-0.16	&	-0.75	&	-0.42	&	-0.18	&	0.08	&	0.53	\\
$\Omega_{22}$	&	1	&	1	&	1	&	1	&	1	&	1	\\
$\theta_1$	&	15.02	&	7.77	&	10.82	&	13.51	&	17.4	&	31.15	\\
$\theta_2$	&	1.72	&	0.17	&	0.75	&	1.3	&	2.2	&	5.64	\\
$V_{11}$	&	266.32	&	60.37	&	116.98	&	182.43	&	302.87	&	970.36	\\
$V_{12}$	&	-4.18	&	-37.77	&	-7.63	&	-2.19	&	0.97	&	19.05	\\
$V_{21}$	&	-4.18	&	-37.77	&	-7.63	&	-2.19	&	0.97	&	19.05	\\
$V_{22}$	&	5.2	&	0.03	&	0.56	&	1.69	&	4.85	&	31.83	\\
$\mu_{K+1}$	&	18.67	&	-16.59	&	8.57	&	18.66	&	28.7	&	54.17	\\
$\tau_{K+1}$	&	-0.79	&	-6.27	&	-1.7	&	-0.64	&	0.16	&	4.05	\\

   \hline
\end{tabular}
\end{table}

\end{document}